\documentclass[prd,preprint,superscriptaddress,tightenlines,nofootinbib, eqsecnum]{revtex4-2}

\usepackage{amsmath}
\usepackage{amsfonts}
\usepackage{amssymb}
\usepackage{bm}
\usepackage[colorlinks]{hyperref}
\usepackage{mathrsfs}
\usepackage{graphicx}
\usepackage{empheq}
\usepackage{ulem}
\usepackage{tensor}
\usepackage[usenames]{color}
\definecolor{darkgreen}{rgb}{0,0.5,0}
\hypersetup{urlcolor=darkgreen}
\usepackage[capitalize]{cleveref}

\allowdisplaybreaks

\DeclareSymbolFontAlphabet{\mathrsfs}{rsfs}
\DeclareMathAlphabet{\mathcal}{OMS}{cmsy}{m}{n}

\newcommand\calO{{\mathcal{O}}}
\newcommand{\dd}{\mathrm{d}}
\newcommand{\di}{\mathrm{i}}
\newcommand{\de}{\mathrm{e}}
\newcommand{\dm}{\mathrm{m}}
\newcommand{\aem}{\alpha_\text{EM}}
\newcommand{\J}{\mu}

\newcommand{\PF}{\underset{B=0}{\text{PF}}}
\newcommand{\go}{\mathfrak{g}}
\newcommand{\dI}{\mathrm{I}}
\newcommand{\dM}{\mathrm{M}}
\newcommand{\dJ}{\mathrm{J}}
\newcommand{\dW}{\mathrm{W}}
\newcommand{\dX}{\mathrm{X}}
\newcommand{\dY}{\mathrm{Y}}
\newcommand{\dZ}{\mathrm{Z}}
\newcommand{\dU}{\mathrm{U}}
\newcommand{\dV}{\mathrm{V}}
\newcommand{\dE}{\mathrm{E}}
\newcommand{\dB}{\mathrm{B}}
\newcommand{\dK}{\mathrm{K}}
\newcommand{\dQ}{\mathrm{Q}}
\newcommand{\dH}{\mathrm{H}}

\newcommand\underrel[2]{\mathrel{\mathop{#2}\limits_{#1}}}

\begin{document}
	
\title{Electromagnetic fields in compact binaries: post-Newtonian wave generation and application to double white dwarfs systems}

\author{Quentin \textsc{Henry}}\email{quentin.henry@aei.mpg.de}
\affiliation{Max Planck Institute for Gravitational Physics\\ (Albert Einstein Institute), D-14476 Potsdam, Germany}

\author{Fran\c{c}ois \textsc{Larrouturou}}\email{francois.larrouturou@desy.de}
\affiliation{Deutsches Elektronen-Synchrotron DESY, Notkestr. 85, 22607 Hamburg, Germany}

\author{Christophe \textsc{Le Poncin-Lafitte}}\email{christophe.leponcin@obspm.fr}
\affiliation{
SYRTE, Observatoire de Paris, Université PSL, CNRS, Sorbonne Université, LNE, \\ 
61 avenue de l’Observatoire, 75014 Paris, France}

\date{\today}

\preprint{DESY-23-074}

\begin{abstract}

The aim of this work is twofold:
(i) to properly define a wave-generation formalism for compact-supported sources embedded in Einstein-Maxwell theory, relying on matched post-Newtonian and multipolar-post-Minkowskian expansions;
(ii) to apply this formalism (which is valid for any type of post-Newtonian sources) to the case of two stars with constant and aligned magnetic dipoles, by computing the fluxes of energy and angular momentum to the next-to-leading order, as well as the gravitational amplitude modes.
Assuming eccentric orbits, we derive the evolution of orbital parameters, as well as the observables of the system, notably the gravitational phase for quasi-circular orbits.
Finally, we give some numerical estimates for the contribution of the magnetic dipoles for some realistic systems.
\end{abstract}

\maketitle

\section{Introduction}
\label{sec_introduction}

High accuracy analytic predictions are a milestone of the signal analysis for gravitational-wave (GW) detectors, notably when it comes to observing the long inspiral phase of coalescing compact objects.
If such objects, typically neutron stars (NS) and white dwarfs (WD), are far from being the main source of events of the current LIGO-Virgo-KAGRA network~\cite{LIGOScientific:2021djp}, they will be a major source for the next generations of detectors, such as the Laser Interferometer Space Antenna (LISA)~\cite{LISA:2017pwj}, or the Einstein Telescope (ET)~\cite{ET:2022}.
Indeed, those instruments will resolve tens of thousands of galactic binaries, composed primarily of WD and/or NS~\cite{Hils:1990,Timpano:2005gm}.
It is thus crucial to provide accurate analytic templates for those sources, notably taking into account physical effects beyond the commonly used spinning point-particle approximation.
Among those effects are for instance the tidal response of the stars, that have been treated to high accuracy in~\cite{Henry:2019xhg,Henry:2020ski}, or the presence of strong magnetic fields, that can be as intense as $10^9\, G$ for white dwarfs and $10^{12}\,G$ for neutron stars~\cite{Ferrario:2006ib,Ferrario:2020}.
Such fields can induce an orbital decay, and shift the frequency of the gravitational wave emitted by the binary~\cite{Bourgoin:2021yzf,Bourgoin:2022ilm,Bourgoin:2022qex,Bourgoin:2022ibr,Carvalho:2022pst,Savalle:2023zbb,Aykroyd:2023cvt}.
The LISA experiment will resolve a large number of double WD~\cite{Korol:2018ulo,Korol:2021pun}, and some of those systems will even be used for the calibration of the instrument~\cite{Stroeer:2006rx,Kupfer:2018jee,Finch:2022prg}. 
It seems therefore important to provide accurate analytic waveform templates that incorporate the electromagnetic structure of WD and NS, which is the purpose of this work.

The templates that will be used for the detection and characterization of the inspiral phase of galactic binaries are mainly based on post-Newtonian results~\cite{Mangiagli:2018kpu}, relying on both weak-field and slow-motion approximations (see \emph{e.g.}~\cite{BlanchetLR,Buonanno:2014aza,Porto:2016pyg} for reviews).
In such a framework, it is customary to distinguish between a ``conservative'' sector, describing the motion of the two objects \emph{via} a set of Noetherian quantities, and a ``dissipative'' one, depicting the fluxes at spatial infinity.
Both sectors are linked by the famous flux-balance equations, that describe the loss of energy $E$ and angular momentum $J^i$ of the system as
\begin{equation}\label{eq_balance_eqs}
\left\langle\frac{\dd E}{\dd t}\right\rangle = - \langle \mathcal{F} \rangle
\qquad\text{and}\qquad
\left\langle\frac{\dd J^i}{\dd t}\right\rangle = - \langle\mathcal{G}^i\rangle\,,
\end{equation}
where $\mathcal{F}$ and $\mathcal{G}^i$ are respectively the energy and angular momentum fluxes, as detected by an observer at spatial infinity and brackets $\langle \cdot \rangle$ denote the usual orbital average.
We let the interested reader refer to~\cite{Blanchet:2018yqa} for the flux-balance equations associated with the other Noetherian quantities, \emph{i.e.} linear momentum and center-of-mass position.
In the case of a binary system bearing electric charges, the left-hand sides of the flux-balance equations have been widely studied, see \emph{e.g.}~\cite{Julie:2017rpw,Khalil:2018aaj,Patil:2020dme,Gupta:2022spq,Martinez:2022vnx,Martinez:2023oga} but, to the best of our knowledge, a proper treatment of the right-hand sides is missing.
Moreover, and as previously argued, going beyond the simple electric charge case will be important for future GW detectors.
In this spirit, we have achieved in~\cite{Henry:2023guc} the construction of the conservative sector including electromagnetic (EM) effects, and computed the Noetherian energy and angular momentum at next-to-leading orders for a binary system of stars bearing magnetic dipoles.
The present work is the continuation of this effort, and we aim at properly defining the corresponding dissipative sector, and build a consistent wave-generation formalism, taking electromagnetic effects into account.

The computation of PN gravitational waveforms relies on matched post-Newtonian and multipolar-post-Minkowskian expansions.
This ``PN-MPM'' framework was developed in~\cite{Blanchet:1985sp,Blanchet:1986dk,Blanchet:1987wq,Blanchet:1992br,Blanchet:1998in} and summarized \emph{e.g.} in~\cite{BlanchetLR}.
It enabled to reach high precision in the analytic determination of the gravitational waveform: the phase is currently known at 4.5PN order~\cite{Blanchet:2023bwj}, the gravitational flux for generic orbits and the dominant gravitational modes, at 4PN order~\cite{Blanchet:2023sbv}, and the other modes, at 3.5PN order~\cite{Faye:2012we,Henry:2022ccf,Henry:2022dzx} (we recall that the $n$PN order corresponds to a $(v/c)^{2n}$ correction beyond the leading order, where $v$ is the typical velocity of the virialized binary system and $c$ the speed of light).
Besides the PN-MPM formalism, two frameworks have also been developed to deal with the wave generation, namely the direct integration of the relaxed equations~\cite{Epstein:1975} and effective field theory~\cite{Goldberger:2004jt,Ross:2012fc}.
They both have reached the 2PN precision for point-particle~\cite{Will:1996zj,Leibovich:2019cxo}.
The aim of the present work is thus to construct a proper PN-MPM formalism including electromagnetic effects, and to apply it to derive the waveform emitted by a binary of stars bearing magnetic dipoles.

The plan of this paper is as follows.
In Sec.~\ref{sec_wave_gen}, we present the construction of the PN-MPM wave-generation formalism including electromagnetic effects, and compute the first non-linearities (the so-called \emph{tails}) in Sec.~\ref{sec_tail}.
Then, we apply our formalism to the case of two spinless stars bearing constant and aligned magnetic moments in Sec.~\ref{sec_toy_model}, and derive the various fluxes, the evolution of orbital parameters for elliptic trajectories, as well as the gravitational phase and modes for quasi-circular orbits.
We also give numerical estimates for those quantities, before concluding in Sec.~\ref{sec_CCL}.
App.~\ref{app_EM_fluxes} presents the construction of the electromagnetic fluxes.
Lengthy expressions are displayed in App.~\ref{app_lengthy_expr}, and stored in an ancillary file~\cite{AncFile}.

\section{Wave-generation formalism}\label{sec_wave_gen}

The aim of this section is to build a wave-generation formalism in the Einstein-Maxwell theory, directly inspired from the usual PN-MPM framework used in GR~\cite{Blanchet:1985sp,Blanchet:1986dk,Blanchet:1987wq,Blanchet:1992br,Blanchet:1998in} (see~\cite{BlanchetLR} for a review).
In such framework, the behavior of the fields at future null infinity is parametrized by some \emph{radiative} multipole moments, from which the fluxes of energy and angular momentum, as well as the gravitational amplitude modes, are derived.
Those radiative moments are linked to the \emph{source} multipole moments, describing the matter distribution, by taking into account the non-linearities arising during the propagation of the waves towards spatial infinity.

As clear from the analysis of the generic structure of the fields in the exterior zone, performed in Sec~\ref{subsec_gen}, three steps are required to construct a proper PN-MPM wave-generation formalism: (i)~expressing the \emph{source} moments in terms of the matter distribution, as done in Sec.~\ref{subsec_source}; (ii)~taking into account the non-linearities \emph{via} an iteration scheme and extracting the \emph{radiative} moments, as discussed in Sec.~\ref{subsec_iteration}; (iii)~relating those moments to the modes and fluxes, as presented in Sec.~\ref{subsec_flux}.

\subsection{Equations of motion}\label{subsec_eoms}

In the following\footnote{The conventions employed throughout this work are as follows: we use a mostly plus signature, the Minkowski metric being $\eta_{\mu\nu} = (-,+,+,+)$; greek letters denote spacetime indices $\mu,\nu,\ldots = (0,1,2,3)$ and latin ones, purely spatial indices $i,j,\ldots = (1,2,3)$; bold font denotes three-dimensional vectors, \emph{e.g.} $\boldsymbol{y}_A = y_A^i$; we use multi-index notations, \emph{i.e.} $\dI_L = \dI_{i_1i_2\ldots i_\ell}$; hats and brackets denote symmetric and trace-free (STF) operators: $\hat{\dI}_L = \dI_{\langle L\rangle} = \underset{L}{\text{STF}}\left[\dI_L\right]$; the d'Alembertian operator is defined with respect to the flat Minkowski metric $\Box \equiv \eta^{\mu\nu}\partial_{\mu\nu} = \Delta - c^{-2}\partial_t^2$; dots and numbers in parenthesis denote time differentiation $A^{(n)} = \dd^nA/\dd t^n$; (anti-)symmetrizations are weighted, \emph{e.g.} $A_{(ij)} = (A_{ij} + A_{ji})/2$.}, we consider a generic matter distribution with compact support in an Einstein-Maxwell framework.
The metric perturbation $h^{\mu\nu} \equiv \sqrt{-g}g^{\mu\nu}-\eta^{\mu\nu}$ and EM field\footnote{Excepted for the numerical applications of Sec.~\ref{sec_toy_model}, we work with the dimensionless field defined in Sec.~III.A of~\cite{Henry:2023guc}, namely $A_\mu \equiv \sqrt{e^2G/c^3\hbar}\,A^\text{dimfull}_\mu$, where $A^\text{dimfull}_\mu$ denotes the usual electromagnetic field.} $A_\mu$ are governed by the coupled fields equations~\cite{Henry:2023guc}
\begin{equation}\label{eq_BoxhA}
\Box h^{\mu\nu} = \frac{16\pi G}{c^4}\,\tau^{\mu\nu}
\qquad\text{and}\qquad
\Box A_\mu = - \frac{4\pi G\aem}{c^4}\,\iota_\mu\,,
\end{equation}
together with the gauge conditions
\begin{equation}\label{eq_gaugecond_hA}
\partial_\nu h^{\mu\nu} = 0
\qquad\text{and}\qquad
\sqrt{-g}\,\nabla^\mu A_\mu = \left( \eta^{\mu\nu}+h^{\mu\nu}\right)\partial_\mu A_\nu = 0\,,
\end{equation}
where $g$ denotes the determinant of $g_{\mu\nu}$, and $\aem = \frac{\mu_0e^2c}{4\pi \hbar} $ is the fine-structure constant, \emph{i.e.} a dimensionless number.
The source terms of the wave equations~\eqref{eq_BoxhA}, $\tau^{\mu\nu}$ and $\iota_\mu$, play the role of the Landau-Lifshitz pseudotensor in GR, and read
\begin{subequations}\label{eq_tau_iota_expl}
\begin{align}
&
\tau^{\mu\nu} \equiv \vert g\vert \,T^{\mu\nu} +  \frac{c^4}{16\pi G}\,\Lambda^{\mu\nu}\,,
\label{eq_tau_expl}\\ 
& 
\iota_\mu \equiv J_\mu  - \frac{c^4}{4\pi G\aem}\, \Phi_\mu\,.
\label{eq_iota_expl}
\end{align}
\end{subequations}
We have separated the sources between the descriptions of the matter distribution, $\{T^{\mu\nu},J_\mu\}$, and the non-linear interaction terms $\{\Lambda^{\mu\nu},\Phi_\mu\}$.
The contribution $\Lambda^{\mu\nu}$ can be further split between the metric self-interaction term $\Lambda^{\mu\nu}_g$ and the interplay between the metric and the EM field, dubbed $\Lambda_\text{EM}^{\mu\nu}$.
The metric self-interaction $\Lambda^{\mu\nu}_g$ is the usual GR one and can found \emph{e.g.} in Eq.~(24) of~\cite{BlanchetLR}.
The two other contributions read explicitly~\cite{Henry:2023guc}
\begin{subequations}\label{eq_LambdaEM_Phi_expl}
\begin{align}
\Lambda^{\mu\nu}_\text{EM} &= \frac{4|g|}{\aem}\left(F^{\mu\lambda}F^\nu_{\ \lambda}-\frac{g^{\mu\nu}}{4}F_{\alpha\beta}F^{\alpha\beta} \right) \,,
\label{eq_LambdaEM_expl}\\
\Phi_\mu &= - h^{\alpha\beta}\partial_{\alpha\beta}A_\mu  - \partial_\mu \go^{\alpha\beta} \,\partial_\alpha A_\beta - \go_{\mu\nu} \go^{\alpha\lambda}\,F_{\alpha\beta}\partial_\lambda\go^{\beta\nu} + \frac{1}{2}\,\go^{\alpha\beta}\go_{\lambda\rho}\,F_{\alpha\mu}\partial_\beta\go^{\lambda\rho}\,,
\label{eq_Phi_expl}
\end{align}
\end{subequations}
where we have introduced the usual field strength $F_{\mu\nu} \equiv \partial_\mu A_\nu - \partial_\nu A_\mu$ together with the ``gothic'' metric $\go^{\mu\nu} \equiv \sqrt{-g}g^{\mu\nu}$.

As for the matter sources $\{T^{\mu\nu},J_\mu\}$, their precise expressions do not play a role in the present section. Indeed our formalism applies to any system with \emph{compact-supported} post-Newtonian matter distribution.
Nevertheless, when applying the techniques developed in the present section to a concrete example in Sec.~\ref{sec_toy_model}, we will need to specify a source modeling.
Following what was done in~\cite{Henry:2023guc}, we will consider a compact binary of spinless point particles, dressed with EM dipoles. Their dynamics is free and will be specified later on. We have
\begin{subequations}\label{eq_Tmunu_Jmu_expl}
\begin{align}
T^{\mu\nu} &= \sum_A \sqrt{\vert g\vert_A}\, u_A^0 v_A^\mu v_A^\nu \bigg(m_A-\frac{1}{2 c^2}\,
\left(F_{\lambda\rho}\right)_A \mathcal{D}^{\lambda\rho}_A\bigg)\,\delta_A\,,
\label{eq_Tmunu_expl}\\
J_\mu &= g_{\mu\nu}\sum_A \partial_\lambda\left( \frac{\mathcal{D}^{\nu\lambda}_A}{u_A^0}\delta_A \right)\,,
\label{eq_Jmu_expl}
\end{align}
\end{subequations}
where $\delta_A\equiv \delta[x^i-y_A^i(t)]$ is the 3-dimensional Dirac delta distribution, locating the matter distribution on the worldline of the particle $A$, $\boldsymbol{y}_A(t)$ (thus the sources are indeed compact supported), $u_A^0 \equiv [-(g_{\mu\nu})_A\,v_A^\mu v_A^\nu/c^2]^{-1/2}$ is the Lorentz factor, $v_A^\mu = (c,v_A^i)$ (with $v_A^i$ the usual 3-dimensional velocity), and the subscript $A$ means that the quantity is regularized in the worldline of particle $A$, using the techniques of~\cite{Blanchet:2000nu}.
The physical quantities describing the source are the individual masses $m_A$ and the EM dipoles, gathered into a Lorentz tensor
\begin{equation}\label{eq_Dmunu_expl}
\mathcal{D}^{\mu\nu}_A \equiv
\begin{pmatrix}
0 & -c\,q^i_A\\
c\,q^i_A & \varepsilon_{ijk}\J^k_A
\end{pmatrix}\,,
\end{equation}
where $\{q_A^i,\J_A^i\}$ are the electric- and magnetic-type dipoles and $\varepsilon_{ijk}$ the Levi-Civita symbol.

Last, but not least, note the difference in structure between the two gauge conditions~\eqref{eq_gaugecond_hA}: while harmonic condition on the metric is linear, the $U(1)$ gauge condition on the EM field has a non-linear sector, involving $h^{\mu\nu}$. This fact will be crucial when explicitly implementing the MPM iteration scheme in Sec.~\ref{subsec_tail_comp}.

\subsection{Generic structure of the fields in the exterior zone}\label{subsec_gen}

The equations of motion~\eqref{eq_BoxhA} are formally solved by the application of a retarded three-dimensional Green function
\begin{subequations}\label{eq_prophA}
\begin{align}
&
h^{\mu\nu} = \frac{16\pi G}{c^4}\,\Box_\text{ret}^{-1}\tau^{\mu\nu}
= -\frac{4G}{c^4}\int\!\!\dd^3\boldsymbol{x}' \, \frac{\tau^{\mu\nu}\left(\boldsymbol{x}',t-\vert \boldsymbol{x}-\boldsymbol{x}'\vert/c\right)}{\vert \boldsymbol{x}-\boldsymbol{x}'\vert}\,,\\
&
A_\mu = 
- \frac{4\pi G\aem}{c^4}\,\Box_\text{ret}^{-1}\iota_\mu
= \frac{G\aem}{c^4}\int\!\!\dd^3\boldsymbol{x}'\, \frac{\iota_\mu\left(\boldsymbol{x}',t-\vert \boldsymbol{x}-\boldsymbol{x}'\vert/c\right)}{\vert \boldsymbol{x}-\boldsymbol{x}'\vert}\,,
\end{align}
\end{subequations}
but those integrals are too complicated to be solved directly.
As we are interested with radiation towards spatial infinity, we will consider the problem in the \emph{exterior} zone, \emph{i.e.} the region of space outside the matter distribution.
In such region of space, both fields $h^{\mu\nu}$ and $A_\mu$ coincide with their multipolar expansions, $\mathcal{M}\left(h^{\mu\nu}\right)$ and $\mathcal{M}\left(A_\mu\right)$ (as those are formal series).
Moreover, we have by construction $T^{\mu\nu} = J_\mu = 0$ in this region.
As multipolar expansions commute with products and derivatives, we then have to solve a simpler, multipolar expanded, version of the equations of motion~\eqref{eq_BoxhA}
\begin{subequations}\label{eq_BoxhA_mult}
\begin{align}
&
\Box\, \mathcal{M}\left(h^{\mu\nu}\right) 
= \mathcal{M}\left(\Box h^{\mu\nu}\right)
= \mathcal{M}\left(\Lambda^{\mu\nu}\right)
= \Lambda^{\mu\nu}\left[\mathcal{M}\left(h^{\mu\nu}\right),\mathcal{M}\left(A_\mu\right)\right]\,,\\
&
\Box\, \mathcal{M}\left(A_\mu\right) 
= \mathcal{M}\left(\Box A_\mu\right)
= \mathcal{M}\left(\Phi_\mu\right)
= \Phi_\mu\left[\mathcal{M}\left(h^{\mu\nu}\right),\mathcal{M}\left(A_\mu\right)\right]\,.
\label{eq_BoxA_mult}
\end{align}
\end{subequations}
Solving the vacuum equations~\eqref{eq_BoxhA_mult} is simpler than computing the integrals~\eqref{eq_prophA}.
Nevertheless, the $r\to \infty$ expansion performed induces spuriously divergent integrals, that have to be regularized.
Let us emphasize that those divergences are not physical, but simply a direct consequence of the formal multipolar expansion.
In order to deal with them, we employ the usual \textit{Hadamard regularization} scheme~\cite{Hadamard1932,Blanchet:2000nu}, introducing an unphysical (length) scale $r_0$, thus solving~\eqref{eq_BoxhA_mult} as\footnote{For high PN computations, it is now customary to use a dimensional regularization scheme instead of a Hadamard one~\cite{Blanchet:2003gy,Blanchet:2005tk}. Nevertheless, for the low PN order derivation of the present work, the Hadamard one is largely sufficient and we let the question of dimensional regularization to future studies.}
\begin{subequations}\label{eq_prophA_mult}
\begin{align}
&
\mathcal{M}\left(h^{\mu\nu}\right) 
= \PF\,\Box_\text{ret}^{-1}\left[\left(\frac{r}{r_0}\right)^B\,\mathcal{M}\left(\Lambda^{\mu\nu}\right)\right] + h_\text{hom}^{\mu\nu}\,,\\
&
\mathcal{M}\left(A_\mu\right) 
= \PF\,\Box_\text{ret}^{-1}\left[\left(\frac{r}{r_0}\right)^B\,\mathcal{M}\left(\Phi_\mu\right) \right] + A^\text{hom}_\mu\,,
\end{align}
\end{subequations}
Note that, together with the regularized propagator acting on the non-linear sources, we have introduced homogeneous solutions, $ h_\text{hom}^{\mu\nu}$ and $A^\text{hom}_\mu$.
Imposing that those homogeneous solutions are regular at spatial infinity (as we recall that we are working in the exterior zone), and the usual \emph{no-incoming radiation} condition (\emph{i.e.} that no wave can ``come from infinity''), they have to be written as
\begin{subequations}\label{eq_hA_FG}
\begin{align}
&
h_\text{hom}^{\mu\nu} = - \frac{4G}{c^4} \sum_{\ell \geq 0} \frac{(-)^\ell}{\ell !}\partial_L \left[\frac{1}{r}\,F_L^{\mu\nu}\left(t-r/c\right)\right]\,,
\label{eq_h_FG}\\
&
A^\text{hom}_\mu = \frac{G\aem}{c^4} \sum_{\ell \geq 0} \frac{(-)^\ell}{\ell !}\partial_L \left[\frac{1}{r}\,G^L_\mu\left(t-r/c\right)\right]\,.
\label{eq_A_FG}
\end{align}
\end{subequations}
where the quantities $F^{\mu\nu}_L$ and $G_\mu^L$ are yet unspecified functions, STF in $L$.
It is easy to see that those homogeneous solutions indeed satisfy the linear wave equations in vacuum $\Box h_\text{hom}^{\mu\nu} = \Box A^\text{hom}_\mu = 0$.

The formal separation of the solution~\eqref{eq_prophA_mult} bears a physical meaning.
On the one hand, the homogeneous solutions $h_\text{hom}^{\mu\nu}$ and $A^\text{hom}_\mu$ represent the linear wave emitted by the source.
As such, they encrypt information on the matter distribution, as presented in Sec.~\ref{subsec_source}.
On the other hand, the regularized propagators acting on the non-linear sources $\Lambda^{\mu\nu}$ and $\Phi_\mu$ describe scattering and diffusion processes happening during the propagation of the wave towards spatial infinity.
They are discussed in Sec.~\ref{subsec_iteration}.

\subsection{Expression of the source moments}\label{subsec_source}

As the source moments encompass the information on the matter system, we need to express the linear (homogeneous) solution in terms of the matter distribution.
The aim is twofold: (i) decomposing the quantities $F_L^{\mu\nu}$ and $G_\mu^L$ entering~\eqref{eq_hA_FG} as simple, irreducible STF \emph{source} (and \emph{gauge}) moments; (ii) matching those moments to the matter distributions $\tau^{\mu\nu}$ and $\iota_\mu$.

The wave equation for $h^{\mu\nu}$ in terms of $\tau^{\mu\nu}$~\eqref{eq_BoxhA} is the same as the one used in pure GR.
Indeed, in the metric sector, the only difference between this work and usual GR computations is the explicit expression of $\tau^{\mu\nu}$ in terms of the fields.
Therefore, the STF reduction and matching of the moments in terms of $\tau^{\mu\nu}$ will be exactly the same as the usual one, derived in~\cite{Blanchet:1998in}.
The linear solution is thus parametrized by six sets STF moments as~\cite{Blanchet:1985sp}
\begin{equation}\label{eq_hhom2hcan}
h^{\mu\nu}_\text{hom} = h^{\mu\nu}_\text{can}\left[\dI_L,\dJ_L\right] + \partial\varphi^{\mu\nu}\left[\dW_L,\dX_L,\dY_L,\dZ_L\right]\,.
\end{equation}
This generic homogeneous solution is split in two sectors.
The \emph{canonical} linear metric $h^{\mu\nu}_\text{can}$ bears the two propagating degrees of freedom of GR, and is written in terms of two sets of \emph{source} moments, $\{\dI_L,\dJ_L\}$~\cite{Sachs:1958zz,Pirani:1964,Thorne:1980ru}. 
Its expression can be found \emph{e.g.} in Eq.~(36) of~\cite{BlanchetLR}.
The second sector, $\partial\varphi^{\mu\nu} \equiv \partial^\mu\varphi^\nu+\partial^\nu\varphi^\mu - \partial_\lambda\varphi^\lambda\,\eta^{\mu\nu}$, is nothing but the generator of a diffeomorphism transformation, parametrized by four sets of \emph{gauge} moments $\{\dW_L,\dX_L,\dY_L,\dZ_L\}$, as explicitly displayed \emph{e.g.} in Eq.~(37) of~\cite{BlanchetLR}.
The matched expressions of those moments in terms of $\tau^{\mu\nu}$ are to be found in Eqs.~(123) and (125) of~\cite{BlanchetLR}.
Note that the monopole $\dI$ and the dipoles $\dI_i$ and $\dJ_i$ are respectively the conserved Arnowitt-Deser-Misner (ADM) mass, linear momentum and angular momentum.
We will operate in the center-of-mass frame, in which $\dI_i = 0$, and denote $\dM = \dI$.

\subsubsection{Irreducible decomposition of the EM field}\label{subsubsec_STFdecomp}

We need to perform the decomposition of the linear EM field~\eqref{eq_A_FG} in irreducible STF tensors, while ensuring that the gauge equation~\eqref{eq_gaugecond_hA} is respected.
Hereafter, for a quantity $Q_L$, we use the shortened notation
\begin{equation}\label{eq_QL_tilde}
\tilde{Q}_L \equiv \frac{1}{r}\,Q_L\left(t-r/c\right)\,.
\end{equation}
Following Eq.~(5.3) of~\cite{Blanchet:1998in}, let us decompose
\begin{equation}
G^L_0 = R_L
\qquad\text{and}\qquad
G^L_i = T^{(+)}_{iL} + \varepsilon^{}_{ai\langle i_\ell}\,T^{(0)}_{L-1\rangle a} + \delta^{}_{i\langle i_\ell}T^{(-)}_{L-1\rangle}\,,
\end{equation}
where
\begin{equation}\label{eq_T_funcG}
T^{(+)}_{iL} = \underset{iL}{\text{STF}}\, G_i^L\,,
\qquad
T^{(0)}_L = \frac{\ell}{\ell+1}\,\varepsilon_{ab\langle i_\ell}^{}\, G_a^{L-1\rangle b}
\qquad\text{and}\qquad
T^{(-)}_{L-1} = \frac{2\ell-1}{2\ell+1}\, G_a^{aL-1}\,.
\end{equation}
Using the property $\hat{\partial}_{iL-1} = \partial_i\hat{\partial}_{L-1}  - \frac{\ell-1}{2\ell-1}\,\delta_{i\langle i_{\ell-1}}\hat{\partial}_{L-2\rangle}\Delta$ together with the homogeneity relation $\Delta \tilde{T}^{(-)}_L = \ddot{\tilde{T}}^{(-)}_L/c^2$, it comes
\begin{subequations}\label{eq_Ahom_RT}
\begin{align}
&
A^\text{hom}_0 = \frac{G\aem}{c^4} \sum_{\ell \geq 0} \frac{(-)^\ell}{\ell !}\partial_L \tilde{R}_L\,,\\
&
A^\text{hom}_i 
= \frac{G\aem}{c^4} \sum_{\ell \geq 0} \frac{(-)^\ell}{\ell !}\Bigg[\frac{\ell\,\partial_{L-1} \ddot{\tilde{T}}^{(-)}_{iL-1}}{(\ell+1)(2\ell+1)c^2} - \ell\,\partial_{L-1} \tilde{T}^{(+)}_{iL-1} + \varepsilon^{}_{ijk}\, \partial_{jL-1} \tilde{T}^{(0)}_{k L-1} - \frac{\partial_{iL}\tilde{T}^{(-)}_L}{\ell+1}\Bigg]\,.
\end{align}
\end{subequations}

Then, we impose the gauge condition
\begin{equation}\label{eq_dmuAmu_Amu}
\go^{\mu\nu}\partial_\mu A_\nu^\text{hom} = \eta^{\mu\nu}\partial_\mu A_\nu^\text{hom} + h^{\mu\nu}\partial_\mu A_\nu^\text{hom} = -\frac{1}{c}\,\dot{A}_0^\text{hom}+\partial_i A_i^\text{hom} + \mathcal{O}(G^2)\,.
\end{equation}
As we are investigating the \emph{linear} solution, we can drop the $\mathcal{O}(G^2)$ remainder for now.
Nevertheless, this remainder will be crucial during the iteration procedure, as explicitly shown in Sec.~\ref{subsubsec_tail_gauge}.
The linear gauge condition thus translates into the relation
\begin{equation}\label{eq_dmuAmu_RT}
\eta^{\mu\nu}\partial_\mu A_\nu^\text{hom} = -\frac{G\aem}{c^4} \sum_{\ell \geq 0} \frac{(-)^\ell}{\ell !}\partial_L \Bigg[ \frac{\dot{\tilde{R}}_L}{c}+\ell\,\tilde{T}^{(+)}_L + \frac{1}{2\ell+1}\frac{\ddot{\tilde{T}}^{(-)}_L}{c^2}\Bigg]=0\,.
\end{equation}
Substituting $T^{(+)}_L$ in terms of $R_L$ and $T^{(-)}_L$ in~\eqref{eq_Ahom_RT}, it comes
\begin{subequations}\label{eq_Ahom_RT_jauge}
\begin{align}
&
A^\text{hom}_0 = \frac{G\aem}{c^4} \sum_{\ell \geq 0} \frac{(-)^\ell}{\ell !}\partial_L \tilde{R}_L\,,\\
&
A^\text{hom}_i = \frac{G\aem}{c^4} \sum_{\ell \geq 0} \frac{(-)^\ell}{\ell !}\Bigg[\frac{ \partial_{L-1}\dot{\tilde{R}}_{iL-1}}{c} +\frac{\partial_{L-1}\ddot{\tilde{T}}^{(-)}_{iL}}{(\ell+1)\,c^2} + \varepsilon^{}_{ijk}\, \partial_{jL-1} \tilde{T}^{(0)}_{k L-1} - \frac{\partial_{iL}\tilde{T}^{(-)}_L}{\ell+1}\Bigg] \,.
\end{align}
\end{subequations}

Next, it is common knowledge that an vector field obeying a $U(1)$ symmetry only propagates two degrees of freedom, whereas the linear solution~\eqref{eq_Ahom_RT_jauge} contains three sets of moments.
Thus, in the same spirit as the splitting of $h_\text{hom}^{\mu\nu}$~\eqref{eq_hhom2hcan}, $A_\mu^\text{hom}$ can be rearranged into a ``canonical'' EM field, parametrized by two sets of STF \emph{source} moments, $\{\dE_L,\dB_L\}$, of respective electric and magnetic parities, and a $U(1)$ generator, parametrized by a set of STF \emph{gauge} moments, $\{\dK_L\}$
\begin{equation}\label{eq_Ahom2Acan}
A^\text{hom}_\mu = A_\mu^\text{can}\left[\dE_L,\dB_L\right] + \partial_\mu \theta\left[\dK_L\right]\,,
\end{equation}
where we pose
\begin{subequations}\label{eq_Amu_funcEBK}
\begin{align}
&
A^\text{can}_0 = -\frac{G\aem}{c^3} \sum_{\ell \geq 0} \frac{(-)^\ell}{\ell !}\partial_L \tilde{\dE}_L\,,\\
&
A^\text{can}_i = -\frac{G\aem}{c^4} \sum_{\ell \geq 1} \frac{(-)^\ell}{\ell !}\bigg[\partial_{L-1} \dot{\tilde{\dE}}_{iL-1} + \frac{\ell}{\ell+1}\,\varepsilon_{ijk}\,\partial_{jL-1}\,\tilde{\dB}_{kL-1}\bigg]\,,\\
&
\theta = -\frac{G\aem}{c^4} \sum_{\ell \geq 0} \frac{(-)^\ell}{\ell !}\partial_L \tilde{\dK}_L\,.
\end{align}
\end{subequations}
The signs and the powers in $c$ have been fixed to recover the usual results for constant charges and dipoles~\cite{jackson1999classical}.
Comparing with Eq.~\eqref{eq_Ahom_RT_jauge}, one can read
\begin{equation}\label{eq_EBK_funcRT}
\dE_L = - \frac{R_L}{c} - \frac{1}{\ell+1}\,\frac{\dot{T}^{(-)}_L}{c^2}\,,
\qquad
\dB_L = - \frac{\ell+1}{\ell} \,T^{(0)}_L
\qquad\text{and}\qquad
\dK_L = \frac{1}{\ell+1}\,T^{(-)}_L\,.
\end{equation}
Let us note that for $\ell = 0$, the linear gauge condition~\eqref{eq_dmuAmu_Amu} simply yields $\dot{\dE} = 0$, which is nothing but the conservation of the electric monopole: the electric charge defined in our formalism is of the ADM type.
In order to reinforce this fact and avoid confusion, we will denote in the following $\dQ = \dE$.

\subsubsection{Matching the source and gauge moments}\label{subsubsec_matching}

Once the EM field has been decomposed in irreducible STF moments, one needs to link those moments to the parameter of the matter distribution, through the powerful \emph{matching} procedure~\cite{Blanchet:1998in}.
This procedure relies on a few hypothesis, presented \emph{e.g.} in Sec.~4.1 of~\cite{BlanchetLR}: (i) the problem can be split into two zones, that partially overlap; (ii) the fields admit a well-defined PN expansion in the near-zone; (iii) the fields admit a well-defined multipolar expansion in the exterior zone.
Under those hypothesis, one can match the two expansions in the overlapping zone, and derive an explicit expression for the moments in terms of the source of the wave equation.

Hopefully, the structure of our problem is identical to the pure GR one, and all the hypothesis are verified.
Indeed, the first hypothesis is satisfied as long as the matter distribution has a compact support, which is one of the assumptions of this work.
The second one has been verified by the study of the conservative sector, presented in~\cite{Henry:2023guc}.
As for the last one, it directly follows from Eq.~\eqref{eq_BoxhA_mult}.
Therefore, one can apply the method exposed in~\cite{Blanchet:1998in,BlanchetLR} to the EM field and derive
\begin{equation}\label{eq_G_expr}
G^L_\mu = \PF \int\!\!\dd^3\boldsymbol{x} \,\left(\frac{r}{r_0}\right)^B\,\hat{x}_L\,\int_{-1}^1\dd z\,\delta_\ell(z)\,\bar{\iota}_\mu\left(\boldsymbol{x},t+zr/c\right)\,,
\end{equation}
where the bar denotes a PN expansion (\emph{i.e.} a formal $c\to \infty$ series), and we have introduced
\begin{equation}\label{eq_deltal_def}
\begin{aligned}
\int_{-1}^1\dd z\,\delta_\ell(z)\,\bar{\iota}_\mu\left(\boldsymbol{x},t+zr/c\right)
&
\equiv 
\frac{(2\ell+1)!!}{2^{\ell+1}\,\ell!}\int_{-1}^1\dd z\,\left(1-z^2\right)^\ell\,\bar{\iota}_\mu\left(\boldsymbol{x},t+zr/c\right)\\
&
= \sum_{k\geq 0}\frac{(2\ell+1)!!}{(2k)!!\,(2\ell+2k+1)!!}\left(\frac{r}{c}\frac{\partial}{\partial t}\right)^{2k}\bar{\iota}_\mu\left(\boldsymbol{x},t\right)\,.
\end{aligned}
\end{equation}
Combining Eqs.~\eqref{eq_EBK_funcRT}, \eqref{eq_T_funcG} and~\eqref{eq_G_expr}, it finally comes
\begin{subequations}\label{eq_EBK_expr}
\begin{align}
&
\dE_L = - \PF \int\!\!\dd^3\boldsymbol{x}\left(\frac{r}{r_0}\right)^B \int_{-1}^1\!\!\dd z\left\lbrace
\delta_\ell(z)\,\hat{x}_L\,\frac{\bar{\iota}_0}{c} + \frac{2\ell+1}{(\ell+1)(2\ell+3)\,c^2}\,\delta_{\ell+1}(z)\,\hat{x}_{iL}\,\dot{\bar{\iota}}_i \right\rbrace\,,\\
&
\dB_L = \varepsilon_{ab\langle i_\ell}\,\PF \int\!\!\dd^3\boldsymbol{x}\left(\frac{r}{r_0}\right)^B \int_{-1}^1\!\!\dd z\,\delta_\ell(z)\,\hat{x}_{L-1\rangle a}\,\bar{\iota}_b\,,\\
&
\dK_L = \frac{2\ell+1}{(\ell+1)(2\ell+3)}\,\PF \int\!\!\dd^3\boldsymbol{x}\left(\frac{r}{r_0}\right)^B \int_{-1}^1\!\!\dd z\,\delta_{\ell+1}(z)\,\hat{x}_{iL}\,\bar{\iota}_i\,.
\end{align}
\end{subequations}
Those expression are in full agreement with the results of~\cite{Damour:1990gj}, that were derived in the framework of non-relativistic electromagnetism (corresponding to setting $\Phi_\mu = 0$).

\subsection{Multipolar-post-Minkowskian iteration scheme and radiative moments}\label{subsec_iteration}

Once the linear solutions $h_\text{hom}^{\mu\nu}$ and $A^\text{hom}_\mu$ are decomposed in irreducible STF moments, one can turn towards the non-linear sectors of the solutions~\eqref{eq_prophA_mult}.
In order to do so, we proceed \emph{via} a post-Minkowskian expansion
\begin{equation}
\mathcal{M}\left(h^{\mu\nu}\right) = \sum_{n\geq 1}G^n h_{(n)}^{\mu\nu}
\qquad\text{and}\qquad
\mathcal{M}\left(A_\mu\right) = \sum_{n\geq 1}G^n A_{(n)\, \mu}\,,
\end{equation} 
where the linear orders $h_{(1)}^{\mu\nu}$ and $A_{(1)\, \mu}$ are naturally given by the homogeneous solutions~\eqref{eq_hhom2hcan} and~\eqref{eq_Ahom2Acan}.
Note that we combine multipolar and post-Minkowskian expansions, hence the name of the iteration scheme.

\subsubsection{Sketch of the iteration procedure}\label{subsubsec_iteration_scheme}

The procedure is conceptually very simple: the vacuum equations~\eqref{eq_BoxhA_mult} are also expanded in a post-Minkowskian fashion, and solved order by order in $G$.
Obviously, we cannot provide here the full solution: for each practical computation, one has to determine by dimensional analysis which interactions will play a role and compute them using techniques developed notably in~\cite{Blanchet:1992br,Blanchet:1997ji,Blanchet:2022vsm,Trestini:2023wwg}.
Note that the main technical difference between our case and the pure GR computations is due to the non-linear nature of the gauge condition for $A_\mu$.
Nevertheless, verifying the gauge condition on the $n$th PM order $A_{(n)\,\mu}$ only requires the knowledge of lower-order solutions $\{h_{(m < n)}^{\mu\nu},A^{}_{(m<n)\,\mu}\}$, already required to compute  $A_{(n)\,\mu}$.
Therefore this non-linear gauge fixing is not a conceptual difficulty by itself.
We provide in Sec.~\ref{sec_tail} a practical example of such a computation, with the determination of the first non-linearities, namely the \emph{tails}.
An interesting fact is that, due to the form of the interaction term $\Lambda_\text{EM}^{\mu\nu}$~\eqref{eq_LambdaEM_expl}, the EM corrections to the gravitational moments $\{\dI_L,\dJ_L\}$ involve at least two EM moments.
For instance, the lowest order EM interactions correcting the mass quadrupole $\dI_{ij}$ are given by $\dQ\times \dE_{ij}$ and $\dE_i\times \dE_j$, which both enter at a relative 2.5PN order (as shown by a rapid dimensional analysis). 

Note also that, when dealing wih pure GR, it is customary to first  let apart the source-to-source interactions and to focus on computing the interactions between gauge and source moments.
Indeed, such interactions can be recast in the form of a physically equivalent description in terms of only two sets of \emph{canonical} moments~\cite{Blanchet:1996wx,Blanchet:2008je,Blanchet:2022vsm}.
As the structure of the linear EM field~\eqref{eq_Ahom2Acan} is very similar to the one of the linear metric~\eqref{eq_hhom2hcan}, it is expected that a similar physical equivalence exists.
Nevertheless, the difference between source and canonical moments enter at relative 2.5PN for the gravitational moments~\cite{Blanchet:1996wx}, and also for the EM ones.
Indeed, rapid dimensional and parity analysis show that the lowest-order gauge corrections to $\dE_i$ (resp. $\dB_i$) are given by the interactions $\dM\times \dK_i$, $\dW\times \dE_i$ and $\dQ\times \dW_i$ (resp. $\dW\times \dB_i$ and $\dK\times\dJ_i$), which are all multiplied by a factor $G/c^5$.
Such precision is beyond the scope of this work, and the gauge moment will play no role in the practical computations of Sec.~\ref{sec_toy_model}.
We thus let the proper definition of \emph{canonical} EM moments to future works.

\subsubsection{Asymptotic structure of the fields}\label{subsubsec_asympt_structure}

Once multipolar-post-Minkowskian expanded metric and EM fields have been constructed up to the required order, one has to extract the \emph{radiative} moments.
Again, the procedure for the metric is similar to the pure GR case, and we let the reader refer to Sec.~3 of~\cite{BlanchetLR} for details.
We will here concentrate on the EM case, and extract its leading behavior at future null infinity $r \to \infty$, with $u = t-r/c$ constant.
Using the relation
\begin{equation}
\partial_L \bigg[\frac{F(t-r/c)}{r}\bigg] =\frac{(-)^\ell\,n_L}{c^\ell\,r}\, F^{(\ell)}(t-r/c) + \mathcal{O}\left(\frac{1}{r^2}\right)\,.
\end{equation}
the behavior of the linear solution~\eqref{eq_Amu_funcEBK} is easily extracted
\begin{subequations}\label{eq_Alin_asympt}
\begin{align}
&
A_0^\text{lin} = - \frac{G\aem}{c^3\,r} \sum_\ell \frac{1}{c^\ell\,\ell !}n_L E_{L}^{(\ell)}+ \mathcal{O}\left(\frac{1}{r^2}\right)\,,\\
&
A_i^\text{lin} = \frac{G\aem}{c^3\,r} \sum_\ell \frac{1}{c^\ell\,\ell !}\bigg[n_{L-1} E_{iL-1}^{(\ell)} - \frac{\ell}{c(\ell+1)}\,\varepsilon_{iab}n_{aL-1}B_{bL-1}^{(\ell)}\bigg]+ \mathcal{O}\left(\frac{1}{r^2}\right)\,,
\end{align}
\end{subequations}
where the moments are evaluated at $u$.
We recall that the gauge moments $\{\dK_L\}$ represent a $U(1)$ symmetry and, as such, play no physical role.
If the linear solution is perfectly regular at infinity, this will not be the case of the non-linear interactions, that will bear (powers of) logarithms.
Indeed, the most generic structure of a MPM iterated solution at future null infinity is of the type~\cite{BlanchetLR}
\begin{equation}
A_\mu  \sim \sum_{\ell,p} \frac{\ln^pr}{r}\,n_L\,\alpha_{\ell,p}(u) + o\left(\frac{1}{r^2}\right)\,,
\end{equation}
where $p \geq 0$ and the ``small remainder'' translates the polylogarithmic behavior at lower orders.
In order to properly define radiative moments, we need to eliminate those logarithms.

\subsubsection{Radiative moments}\label{subsubsec_radiative_moments}

In pure GR, the study of the asymptotic structure of the space-time has revealed that this polylogarithmic behavior is an artifact of the harmonic coordinates, and there exist ``radiative'' coordinate systems in which the metric is free of it~\cite{Penrose:1962ij,Penrose:1965am}.
The most famous ones are the Bondi coordinates system~\cite{Bondi:1962px,Sachs:1962wk} and the Newman-Unti one~\cite{Newman:1963ugj}, but the class of radiative coordinates is quite large~\cite{Papapetrou:1969,MadoreI,MadoreII}.
Naturally, there exist generic constructions relating harmonic and Bondi-Newman-Unti coordinate systems exist~\cite{Blanchet:1986dk,Blanchet:2020ngx,Blanchet:2023pce} (see~\cite{Trestini:2022tot} for a practical implementation at the state-of-the art 4PN precision).

Our claim here is that there exists a radiative coordinate system $(T,X^i)$ in which both the metric and the EM field are free of polylogarithmic structure.
\begin{equation}\label{eq_HA_rad_genericstructure}
H^{\mu\nu} \sim \sum_{\ell} \frac{N_L}{R}\,H_L(U) + \mathcal{O}\left(\frac{1}{R^2}\right)
\qquad\text{and}\qquad
\mathcal{A}_\mu \sim \sum_{\ell} \frac{N_L}{R}\,A_L(U) + \mathcal{O}\left(\frac{1}{R^2}\right)\,,
\end{equation}
with naturally $U = T-R/c$.
Such claim is corroborated by the analysis of the Sec.~\ref{sec_tail}, were we show that, under the coordinate change~\eqref{eq_coordchangerad2harm} both the metric and the EM field takes the form~\eqref{eq_HA_rad_genericstructure}.

In such coordinate system, one can project the metric $H^{\mu\nu}$ in a traceless and transverse (TT) gauge~\cite{Thorne:1980ru} $H_{0\mu}^\text{TT} = N_i H_{ij}^\text{TT}= H_{ii}^\text{TT}=0$ as
\begin{equation}\label{eq_HijTT}
H_{ij}^\text{TT} \equiv - \frac{4G}{c^2\,R}\,\mathcal{P}^\text{TT}_{ijab} \sum_\ell \frac{1}{c^\ell\,\ell !}\bigg[N_{L-2} \dU_{abL-2} - \frac{2\ell}{c(\ell+1)}\,N_{cL-2}\varepsilon_{cd(a}\dV_{b)dL-2}\bigg]
+ \mathcal{O}\left(\frac{1}{R^2}\right)\,,
\end{equation}
where we introduced the TT projector $\mathcal{P}^\text{TT}_{ijab} = \perp_{i(a} \perp_{b)j} - \frac{1}{2}\perp_{ij}\perp_{ab}$ with $\perp_{ij} = \delta_{ij} - N_{ij}$.
The two sets of moments $\{\dU_L,\dV_L\}$ entering the asymptotic metric are naturally dubbed mass and current \emph{radiative} moments.

As for the EM field, let us project it on a transverse basis, defined by $A_0^\text{T} = N_iA_i^\text{T} = 0$, as
\begin{equation}\label{eq_AiT}
A_i^\text{T} = \frac{G\aem}{c^3\,R} \perp_{ij}
\sum_\ell \frac{1}{c^\ell\,\ell !}\bigg[N_{L-1} \dQ_{jL-1} - \frac{\ell}{c(\ell+1)}\,\varepsilon_{jab}N_{aL-1}\dH_{bL-1}\bigg]
+ \mathcal{O}\left(\frac{1}{R^2}\right)\,,
\end{equation}
where similarly, we have introduced two sets of \emph{radiative} electric and magnetic moments, $\{\dQ_L, \dH_L\}$.
Looking back at Eq.~\eqref{eq_Alin_asympt}, it comes as expected
\begin{equation}\label{eq_QLHL_ELBL}
\dQ_L = \dE_L^{(\ell)} + \mathcal{O}(G)
\qquad\text{and}\qquad
\dH_L = \dB_L^{(\ell)} + \mathcal{O}(G)\,,
\end{equation}
where the remainder encrypts all non-linearities, and notably the tails computed in Sec.~\ref{sec_tail}.

\subsection{Computation of the modes and fluxes}\label{subsec_flux}

\subsubsection{Gravitational sector}\label{subsubsec_flux_grav}

The detailed derivation of the energy and angular momentum fluxes associated with the asymptotic metric~\eqref{eq_HijTT} can be found in~\cite{Thorne:1980ru}, and they read
\begin{subequations}\label{eq_flux_FGi_grav}
\begin{align}
&
\mathcal{F}_\text{GW} =
\sum_{\ell\geq 2} \frac{G}{c^{2\ell+1}} \frac{(\ell+1)(\ell+2)}{(\ell-1)\ell\,\ell !(2\ell+1)!!}\bigg[\dU_L^{(1)}\dU_L^{(1)} + \frac{4\ell^2}{c^2(\ell+1)^2}\,\dV_L^{(1)}\dV_L^{(1)}\bigg]
\label{eq_flux_F_grav}\,,\\
&
\mathcal{G}^i_\text{GW} =
\varepsilon_{iab}\,\sum_{\ell\geq 2} \frac{G}{c^{2\ell+1}} \frac{(\ell+1)(\ell+2)}{(\ell-1)\,\ell !(2\ell+1)!!}\bigg[\dU^{}_{aL-1}\dU_{bL-1}^{(1)} + \frac{4\ell^2}{c^2(\ell+1)^2}\,\dV^{}_{aL-1}\dV_{bL-1}^{(1)}\bigg]
\,.
\label{eq_flux_Gi_grav}
\end{align}
\end{subequations}
From the asymptotic metric~\eqref{eq_HijTT}, one can also extract the observable gravitational modes, $h_{\ell \dm}$, by projecting $H_{ij}^\text{TT}$ onto the basis of polarization $\{+,\times\}$ and on spin-weighted spherical harmonics, $Y_{-2}^{\ell \dm}$ (following the conventions of~\cite{Blanchet:2008je,Faye:2012we}) as
\begin{equation}\label{eq_mode_hlm_def}
h_+ - \di h_\times
= \sum_{\ell \geq 2}\sum_{\dm=-\ell}^{\ell}h_{\ell \dm} \,Y_{-2}^{\ell \dm}\,.
\end{equation}
In the particular case of planar orbits, as will be considered in Sec~\ref{sec_toy_model}, the gravitational modes are linked to the radiative moments by the relations~\cite{Faye:2012we}
\begin{subequations}\label{eq_mode_hlm2ULVL}
\begin{align}
&
h_{\ell \dm} = - \frac{2G}{Rc^{\ell+2}\,\ell!}\sqrt{\frac{(\ell+1)(\ell+2)}{\ell(\ell-1)}}\,\alpha_L^{\ell \dm}\,\dU_L\,,
& & 
\text{if}\ \ell + \dm\ \text{is even}\,,\\
&
h_{\ell \dm} = - \frac{4G\,\di}{Rc^{\ell+3}\,\ell!}\sqrt{\frac{\ell(\ell+2)}{(\ell-1)(\ell+1)}}\,\alpha_L^{\ell \dm}\,\dV_L\,,
& & 
\text{if}\ \ell + \dm\ \text{is odd}\,.
\end{align}
\end{subequations}
Introducing a fixed orthonormal basis $(\boldsymbol{n}_0,\boldsymbol{\lambda}_0,\boldsymbol{l})$ where $\boldsymbol{l}$ is normal to the orbital plane, together with $\boldsymbol{\mathfrak{m}}_0 = (\boldsymbol{n}_0+\di \boldsymbol{\lambda}_0)/\sqrt{2}$, the projector $\alpha_L^{\ell \dm}$ is explicitly given by
\begin{equation}\label{eq_alphalmL_def}
\alpha_L^{\ell \dm} = \frac{(-)^\dm\,2^\frac{2+\vert\dm\vert}{2}\sqrt{\pi}\,\ell !}{\sqrt{(2\ell+1) (\ell+\dm)!\,(\ell-\dm)!}}\,\overline{\mathfrak{m}}_0^{\langle M}l^{L-M\rangle}\,,
\end{equation}
where the overbar denotes complex conjugation.

\subsubsection{Electromagnetic sector}\label{subsubsec_flux_EM}

As discussed in App.~\ref{app_EM_fluxes_gen}, the energy loss can be expressed by the angular average flux of the Landau-Lifshitz pseudotensor $\tau^{\mu\nu}$. At leading order in $1/R$, the EM sector of this tensor reduces to the usual Poynting vector $\Pi_i = c\,F^\text{T}_{ij}F^\text{T}_{j0}$. The EM energy flux then reads
\begin{equation}\label{eq_flux_F_EM_def}
\mathcal{F}_\text{EM} 
= \frac{c^4}{4\pi G\,\aem}
\lim_{R\to\infty}\left[\int\!\!\dd^2\Omega\,R^2\,N_a\Pi_a\right]
= \frac{c^3}{4\pi G\,\aem}
\lim_{R\to\infty}\left[\int\!\!\dd^2\Omega\,R^2\,\dot{A}_i^\text{T}\dot{A}_i^\text{T}\right]\,,
\end{equation}
where we used the relation $\partial_j A^\text{T}_i = - N_j\dot{A}^\text{T}_i/c + \mathcal{O}(1/R)$ together with the transversity.
Plugging the asymptotic expression for the EM field~\eqref{eq_AiT}, and performing the angular integrals with the help of the formula~\eqref{eq_app_angular_perp}, it comes
\begin{equation}\label{eq_flux_F_EM}
\mathcal{F}_\text{EM} = \sum_{\ell\geq 1} \frac{G \aem}{c^{3+2\ell}}\frac{\ell+1}{\ell\,\ell !(2\ell+1)!!}\bigg[\dQ_L^{(1)}\dQ_L^{(1)} + \frac{\ell^2}{c^2(\ell+1)^2}\,\dH_L^{(1)}\dH_L^{(1)}\bigg]\,.
\end{equation}
This result is consistent with the computation of~\cite{Ross:2012fc}, and the canonical Larmor formula~\cite{jackson1999classical} is recovered in the case of a simple electric dipole.
Moreover, to emphasize the gauge invariance of the flux, a derivation relaxing the transverse gauge condition is presented in App.~\ref{app_EM_fluxes_F} and fully agree with our result~\eqref{eq_flux_F_EM}.

As for the angular momentum flux, it requires more care, just as in the gravitational case (see \emph{e.g.}~\cite{Bonga:2018gzr,Blanchet:2018yqa,Ashtekar:2019rpv,Veneziano:2022zwh,Manohar:2022dea,Riva:2023xxm} for subtleties linked to the definition of an angular momentum flux in GR).
As discussed in App.~\ref{app_EM_fluxes_gen}, it is also defined with respect to the EM sector of the Landau-Lifshitz pseudotensor~\eqref{eq_LambdaEM_expl} as
\begin{equation}\label{eq_flux_Gi_EM_def}
\begin{aligned}
\mathcal{G}_\text{EM}^i 
&
=\frac{c^4}{16\pi G}\,\lim_{R\to \infty}\left[\varepsilon_{iab}\int\!\!\dd^2\Omega\,R^3 N_{ac}\,\Lambda_\text{EM}^{bc}\right]\\
&
= \frac{c^4}{4\pi G\aem}\,\lim_{R\to \infty}\left[\varepsilon_{iab}\int\!\!\dd^2\Omega \,R^3\,N_{ac}\,\bigg(F_{bd}F_{cd} - F_{0b}F_{0c}\bigg)\right]\,.
\end{aligned}
\end{equation}
But, as clear from the $R^3$ factor, its evaluation requires the knowledge of the sub-dominant $\mathcal{O}(1/R^2)$ order in the asymptotic expansion of the field~\eqref{eq_AiT}.
This is quite similar to what is happening in pure GR, see \emph{e.g.}~\cite{Thorne:1980ru}, and calls for a full BMS-type analysis of the asymptotic structure of the fields.
Instead of performing such tedious computation, we follow the spirit of~\cite{Blanchet:2018yqa} and provide in App.~\ref{app_EM_fluxes_G} a derivation leading to the result
\begin{equation}\label{eq_flux_Gi_EM}
\mathcal{G}^i_\text{EM} = \varepsilon_{iab}\,\sum_{\ell\geq 1} \frac{G \aem}{c^{3+2\ell}}\frac{\ell+1}{\ell !(2\ell+1)!!}\bigg[\dQ_{aL-1}^{}\dQ_{bL-1}^{(1)} + \frac{\ell^2}{c^2(\ell+1)^2}\,\dH_{aL-1}^{}\dH_{bL-1}^{(1)}\bigg]\,.
\end{equation}
Note that, in the case of a rotating electric dipole, we recover the results of~\cite{PhysRevA.105.023511,Landau:1975pou}.

Together with the gravitational fluxes~\eqref{eq_flux_FGi_grav}, those fluxes enter the balance equations for the energy and angular momentum~\eqref{eq_balance_eqs}, that allow to derive the secular evolution of orbital quantities, see Sec.~\ref{subsec_elliptic}.
In the case of quasi-circular orbits, they are key ingredient to derive the gravitational phase $\phi = \int\!\!\dd t\, \omega$ (where $\omega$ is the gravitational frequency), as presented in Sec.~\ref{subsec_circ}.
Interestingly, the two fluxes~\eqref{eq_flux_F_EM} and~\eqref{eq_flux_Gi_EM} start formally at the same order than the gravitational ones, namely 2.5PN.
This is a strong difference with the results of the conservative sector, where the EM effects enter formally as 1PN corrections to the Noetherian quantities~\cite{Henry:2023guc}.

\section{Computation of the leading tail effects}\label{sec_tail}

The leading corrections to the fluxes derived in the previous section are the so-called \emph{tail} interactions.
Those interactions depict the scattering of a propagating moment ($\dI_L$, $\dJ_L$, $\dE_L$ or $\dB_L$) onto the static structure of the space-time, described by the constant (ADM) mass $\dM$.
The gravitational tails $\dM \times \dI_L$ and $\dM \times \dJ_L$ are known since a long time~\cite{Blanchet:1989ki,Blanchet:1992br,Blanchet:1993ec,Blanchet:1995fr} (see also~\cite{Poisson:1993vp} for a derivation using Schwarzschild coordinates, and~\cite{Trestini:2022tot} for a computation in a radiative coordinate system), and we will not redo it here.
Instead we will focus on the leading EM tails, namely the interaction of the EM $\ell = 1$ moments $\dE_i$ and $\dB_i$ with the ADM mass $\dM$, and will explicitly apply the MPM iteration scheme pictured in Sec.~\ref{subsubsec_iteration_scheme} to those interactions.\footnote{Note that the interaction between the constant charge $\dQ$ and the gravitational dipole $\dI_i$ formally enters at the same order than the tail $\dM\times \dE_i$. Nevertheless, we work in the center-of-mass frame, in which $\dI_i = 0$, so we will discard this interaction.}

\subsection{The electro-magnetic tail effect}\label{subsec_tail_comp}

The MPM iteration method~\cite{Blanchet:1992br} is a two-steps procedure, described in details \emph{e.g.} in Sec.~2 of~\cite{BlanchetLR}.
First, one computes a \emph{particular} solution of the sourced wave equation, by applying the Hadamard-regularized propagator.
But, due to the regularization procedure, this particular solution may not obey the gauge condition, even if the source does\footnote{Note that a similar departure from gauge can also happens in the EFT framework, where the computation has to be corrected by a removal of \emph{anomalies}~\cite{Almeida:2023yia}.} and one needs to correct for it.
Therefore, the second step is to add a proper \emph{homogeneous} solution that corrects for the violation of the gauge condition.
The sum of those two solutions is the correct value of the field, from which the radiative moments can be extracted (after the suitable coordinate change, as explained in Sec.~\ref{subsubsec_radiative_moments}).

\subsubsection{Computation of the particular solution}\label{subsubsec_tail_part}

In what follows, we aim to compute the quadratic interactions $\dM\times \dE_i$ and $\dM\times\dB_i$, by solving the PM-expanded vacuum equation~\eqref{eq_BoxA_mult}.
The first step is naturally to construct the non-linear source $\mathcal{M}(\Phi_\mu)$, by injecting the MPM expansions of the fields into the expression of $\Phi_\mu$.
Selecting only the $\dM$ sector of the linear metric~\eqref{eq_hhom2hcan} and the appropriate sectors of the linear EM field~\eqref{eq_Amu_funcEBK}, we need to consider
\begin{subequations}
\begin{align}
&
h^{00}_\dM = -\frac{4G\dM}{c^2\,r}\,,
& &
 h^{0i}_\dM = 0\,,
\qquad\qquad\qquad\quad h^{ij}_\dM = 0\,.\\
&
A_0^{\dE_i} = \frac{G\aem}{c^3}\, \partial_i \tilde{\dE}_i  \,,
& & 
A_i^{\dE_i} = \frac{G\aem}{c^4} \,\dot{\tilde{\dE}}_i\,,\\
&
A_0^{\dB_i} = 0 \,,
& & 
A_i^{\dB_i} = \frac{G\aem}{2c^4}  \partial_j\tilde{\dB}_{i\vert j}\,,
\end{align}
\end{subequations}
where we recall that we employ the shortcut~\eqref{eq_QL_tilde}, and have used the notation of~\cite{Henry:2021cek}, which amounts here to $\dB_{i\vert j} \equiv \varepsilon_{ijk}\dB_k$.
Those MPM-expansions are injected into the non-linear source term~\eqref{eq_Phi_expl} developed to quadratic order
\begin{equation}\label{eq_Phi_quadratic}
\Phi_\mu =
- h^{\alpha\beta}\partial_{\alpha\beta}A_\mu - \partial_\mu h^{\alpha\beta}\partial_\alpha A_\beta - F_{\alpha\beta}\partial^\alpha h^\beta_\mu - \frac{1}{2}\,F_{\mu\alpha}\partial^\alpha h+ \calO(h^2A)\,,
\end{equation}
where indices are operated with the Minkowski metric and $h = h^\mu_\mu$ is the trace of $h^{\mu\nu}$.
It is easy to see that the equations we need to solve are of the form
\begin{equation}
\Box A_\mu = \sum_{p,\ell} \frac{\hat{n}_L}{r^p}\,f_p^L\left(t- \frac{r}{c}\right)\,,
\end{equation}
and do not involve logarithmic dependencies, so we can apply the formulas displayed in App.~1 of~\cite{Blanchet:1997ji} to compute the \emph{particular} solutions
\begin{equation}\label{eq_apart_MEi_MBi_formal}
a_\mu^{\dM\times \dE_i} = \PF \Box_\text{ret}^{-1}  \left[ \left(\frac{r}{r_0}\right)^B\Phi^{\dM\times \dE_i}_\mu\right]
\qquad\text{and}\qquad
a_\mu^{\dM\times \dB_i} = \PF \Box_\text{ret}^{-1}  \left[ \left(\frac{r}{r_0}\right)^B\Phi^{\dM\times \dB_i}_\mu\right]\,,
\end{equation}
which explicitly read\footnote{We recall that numbers in parenthesis denote time differentiation, namely $A^{(n)} = \dd^nA/\dd t^n$.}
\begin{subequations}\label{eq_apart_MEi_MBi}
\begin{align}
a_0^{\dM\times \dE_i}  =
&\,
\frac{4G^2\dM\aem}{c^8}\,n^i\,\bigg[\int_1^\infty\!\!\!\!\!\dd x \,Q_1(x)\,\dE^{(3)}_i\left(t-\frac{xr}{c}\right) 
+ \frac{3c}{4r}\,\dE_i^{(2)} 
+ \frac{c^2}{4r^2}\,\dE_i^{(1)} 
+ \frac{c^3}{4r^3}\,\dE_i\bigg]\,,\\
a_i^{\dM\times \dE_i} =
&\,
- \frac{4G^2\dM\aem}{c^8}\bigg[\int_1^\infty\!\!\!\!\!\dd x \,Q_0(x)\,\dE^{(3)}_i\left(t-\frac{xr}{c}\right) 
- \frac{c}{3r}\,\dE_i^{(2)}
- \frac{c^2}{3r^2}\,\dE_i^{(1)} \\
& \nonumber \hspace{3cm}
+ \frac{c\,\hat{n}^{ij}}{8r}\,\dE_j^{(2)} 
+ \frac{c^2\,\hat{n}^{ij}}{8r^2}\,\dE_j^{(1)}\bigg]\,,\\
a_0^{\dM\times \dB_i}  =
&\, 0\,,\\
a_i^{\dM\times \dB_i}  =
&\, 
\frac{2G^2\dM\aem}{c^8}\frac{n^j}{r}\bigg[\int_1^\infty\!\!\!\!\!\dd x \,Q_1(x)\, \dB_{i\vert j}^{(3)}\left(t-\frac{xr}{c}\right)
+\frac{5}{8}\,\dB^{(2)}_{i\vert j} 
-\frac{c}{8r}\, \dB^{(1)}_{i\vert j}
-\frac{c^2}{8r^2}\,\dB_{i\vert j}\bigg]\,.
\end{align}
\end{subequations}
where the Legendre functions $Q_\ell$ are defined with a branch cut on $(-\infty,1)$ as
\begin{equation}
Q_\ell(x) = \frac{1}{2}\,P_\ell(x) \ln\bigg(\frac{x+1}{x-1}\bigg) - \sum_{j=1}^\ell \frac{1}{j}\,P_{\ell-j}(x)P_{j-1}(x)\,,
\end{equation}
with $P_\ell$ the Legendre polynomials.

\subsubsection{Implementation of the gauge condition}\label{subsubsec_tail_gauge}

Due to the presence of the regularization kernel $(r/r_0)^B$ inside the Green function~\eqref{eq_apart_MEi_MBi_formal}, the solution $a_\mu$~\eqref{eq_apart_MEi_MBi} has \emph{a priori} no reason to verify the gauge condition, as discussed in Sec.~2 of~\cite{BlanchetLR}. 
We thus need to verify it and to correct it, if needed, by adding a suitable homogeneous solution. 

Using relations among the Legendre functions, it comes
\begin{equation}
\begin{aligned}
\partial_i a_i^{\dM\times\dE_i}-\frac{1}{c}\,\partial_t a_0^{\dM\times\dE_i}
&
= - \frac{4G^2\dM\aem}{c^7}\frac{n^i}{r^2}\,\bigg[\dE_i^{(2)} + \frac{c}{r}\,\dE_i^{(1)}\bigg]\,,\\
\partial_i a_i^{\dM\times\dB_i}-\frac{1}{c}\,\partial_t a_0^{\dM\times\dB_i} 
&
= 0\,.
\end{aligned}
\end{equation}
On the other hand, the non-linear part of the gauge relation~\eqref{eq_gaugecond_hA} yields
\begin{equation}
h^{\mu\nu}_\dM\, \partial_\mu A_\nu^{\dE_i} =\frac{1}{c}\,h^{00}_\dM\, \partial_t A_0^{\dE_i} = \frac{4G^2\dM\aem}{c^7}\frac{n^i}{r^2}\,\bigg[\dE_i^{(2)} + \frac{c}{r}\,\dE_i^{(1)}\bigg]\,,
\qquad
h^{\mu\nu}_\dM\, \partial_\mu A_\nu^{\dB_i} = 0\,.
\end{equation}
So the particular solutions~\eqref{eq_apart_MEi_MBi} already satisfy the gauge condition, no further work is needed. 
Note that taking into account the non-linear part of the gauge relation was crucial here.
But, importantly, we cannot generalize this fact, and a gauge-fixing procedure, such as the one described in Sec.~2 of~\cite{BlanchetLR} is expected to be vital for higher-order computations.

The first PM iteration of the EM field for the tail interaction thus reads
\begin{subequations}\label{eq_Atail_MEi_MBi}
\begin{align}
A_0^{\dM\times \dE_i}  =
&\,
\frac{4G^2\dM\aem}{c^8}\,n^i\,\bigg[\int_1^\infty\!\!\!\!\!\dd x \,Q_1(x)\,\dE^{(3)}_i\left(t-\frac{xr}{c}\right) 
+ \frac{3c}{4r}\,\dE_i^{(2)} 
+ \frac{c^2}{4r^2}\,\dE_i^{(1)} 
+ \frac{c^3}{4r^3}\,\dE_i\bigg]\,,\\
A_i^{\dM\times \dE_i} =
&\,
- \frac{4G^2\dM\aem}{c^8}\bigg[\int_1^\infty\!\!\!\!\!\dd x \,Q_0(x)\,\dE^{(3)}_i\left(t-\frac{xr}{c}\right) 
- \frac{c}{3r}\,\dE_i^{(2)}
- \frac{c^2}{3r^2}\,\dE_i^{(1)} \\
& \nonumber \hspace{3cm}
+ \frac{c\,\hat{n}^{ij}}{8r}\,\dE_j^{(2)} 
+ \frac{c^2\,\hat{n}^{ij}}{8r^2}\,\dE_j^{(1)}\bigg]\,,\\
A_0^{\dM\times \dB_i}  =
&\, 0\,,\\
A_i^{\dM\times \dB_i}  =
&\, 
\frac{2G^2\dM\aem}{c^8}\frac{n^j}{r}\bigg[\int_1^\infty\!\!\!\!\!\dd x \,Q_1(x)\, \dB_{i\vert j}^{(3)}\left(t-\frac{xr}{c}\right)
+\frac{5}{8}\,\dB^{(2)}_{i\vert j} 
-\frac{c}{8r}\, \dB^{(1)}_{i\vert j}
-\frac{c^2}{8r^2}\,\dB_{i\vert j}\bigg]\,.
\end{align}
\end{subequations}

\subsection{Correction to the radiative moments}\label{subsec_tail_rad}

The previous computation~\eqref{eq_Atail_MEi_MBi} gives the first PM correction to the MPM expanded EM field.
Let us now extract the corresponding corrections to the radiative moments, following the procedure described in Sec.~\ref{subsubsec_radiative_moments}.

The first step is to develop the fields at future null infinity, \emph{i.e.} $r\to \infty$ with $u = t-r/c$ constant.
Using for instance Eqs.~(4.5) and~(4.6) of~\cite{Trestini:2022tot}, one expands the Legendre functions as
\begin{equation}\label{eq_exp_Ql}
\int_1^\infty\!\!\!\!\!\dd x \,Q_\ell(x)F\bigg(t-\frac{xr}{c}\bigg)
= - \frac{c}{2r}\int_0^\infty\!\!\!\!\!\dd \tau\bigg[\ln\left(\frac{c\tau}{2r}\right) +2 \mathcal{H}_\ell \bigg] F\big(u-\tau\big)+ \mathcal{O}\left(\frac{\ln r}{r^2}\right)\,,
\end{equation}
where $\mathcal{H}_\ell = \sum_{j=1}^\ell 1/j$ is the harmonic number.\footnote{Note that the coefficient in front of the logarithm does not depend on the value of the index of the Legendre function. This fact will be crucial when applying the coordinate change~\eqref{eq_coordchangerad2harm} to remove the $\ln(r)$ dependency.}
Therefore, the asymptotic expansion of the previous result~\eqref{eq_Atail_MEi_MBi} reads
\begin{subequations}\label{eq_A_coordharm_infty}
\begin{align}
&
A_0^{\dM\times \dE_i} = - \frac{2G^2\dM\aem\,n_i}{c^7 \,r}\int_0^\infty\!\!\!\!\!\dd\tau\bigg[\ln\left(\frac{c\tau}{2r}\right) + \frac{1}{2}\bigg] \dE_i^{(3)}\big(u-\tau\big)
+ \mathcal{O}\left(\frac{\ln r}{r^2}\right)\,,\\
&
A_i^{\dM\times \dE_i} = \frac{2G^2\dM\aem}{c^7 \,r}\Bigg\lbrace\int_0^\infty\!\!\!\!\!\dd\tau\bigg[\ln\left(\frac{c\tau}{2r}\right) + \frac{3}{4}\bigg] \dE_i^{(3)}\big(u-\tau\big) -\frac{n^i}{4}n_j\dE^{(2)}_j(u)\Bigg\rbrace
+ \mathcal{O}\left(\frac{\ln r}{r^2}\right)\,,\\
&
A_i^{\dM\times \dB_i} = - \frac{G^2\dM\aem\,n_j}{c^8 \,r}\int_0^\infty\!\!\!\!\!\dd\tau\bigg[\ln\left(\frac{c\tau}{2r}\right) + \frac{3}{4}\bigg] \dB_{i\vert j}^{(3)}\big(u-\tau\big)
+ \mathcal{O}\left(\frac{\ln r}{r^2}\right)\,,
\end{align}
\end{subequations}
and we recall that $A_0^{\dM\times \dB_i} = 0$.

Let us now perform a change of coordinate system, to move from the current harmonic one to a radiative one, where the asymptotic expansion of the metric is free of polylogarithmic structure.
Such transformation is well known (see \emph{e.g.} the detailed discussion in~\cite{Trestini:2022tot}) and reads at lowest order
\begin{equation}\label{eq_coordchangerad2harm}
T = t - \frac{2G\dM}{c^3}\,\ln\left(\frac{r}{c\,b_0}\right) + \mathcal{O}\left(G^2\right)\,,
\qquad
X^i = x^i+ \mathcal{O}\left(G^2\right)\,.
\end{equation}
Here, $b_0$ is an (unphysical) constant linked to the time origin in the new coordinate system, and will disappear from all observable quantities.
Applying this coordinate change to the asymptotic expansion of the linear metric~\eqref{eq_Alin_asympt}, and combining it with~\eqref{eq_A_coordharm_infty}, it turns out that the change of coordinate system simply amounts to the replacement $\ln(r)\to \ln(c\,b_0)$, just as in the GR case.
\begin{subequations}\label{eq_A_coordrad_infty}
\begin{align}
&
\mathcal{A}_0^{\dM\times \dE_i} = - \frac{2G^2\dM\aem\,N_i}{c^7 \,R}\int_0^\infty\!\!\!\!\!\dd\tau\bigg[\ln\left(\frac{\tau}{2b_0}\right) + \frac{1}{2}\bigg] \dE_i^{(3)}\big(U-\tau\big)
+ \mathcal{O}\left(\frac{\ln R}{R^2}\right)\,,\\
&
\mathcal{A}_i^{\dM\times \dE_i} = \frac{2G^2\dM\aem}{c^7 \,R}\Bigg\lbrace \int_0^\infty\!\!\!\!\!\dd\tau\bigg[\ln\left(\frac{\tau}{2b_0}\right) + \frac{3}{4}\bigg] \dE_i^{(3)}\big(U-\tau\big) - \frac{N^i}{4}\,N_j\dE_j^{(2)}\big(U\big)\Bigg\rbrace
+ \mathcal{O}\left(\frac{\ln R}{R^2}\right)\,,\\
&
\mathcal{A}_i^{\dM\times \dB_i} = -\frac{G^2\dM\aem\,N_j}{c^8 \,R}\int_0^\infty\!\!\!\!\!\dd\tau\bigg[\ln\left(\frac{\tau}{2b_0}\right) + \frac{3}{4}\bigg] \dB_{i\vert j}^{(3)}\big(U-\tau\big)
+ \mathcal{O}\left(\frac{\ln R}{R^2}\right)\,.
\end{align}
\end{subequations}
Therefore, the coordinate change~\eqref{eq_coordchangerad2harm} removes the logarithmic structure present in the asymptotic expansions of both the metric and the EM field.
Can we expect this simultaneous cancellation to hold at higher orders?
For a tail with generic multipolarity, $\dM \times \dE_L$ or $\dM\times \dB_L$, the property of the asymptotic expansion of $Q_\ell$~\eqref{eq_exp_Ql} leads us to think that it will be the case.
Indeed, in GR, the single coordinate change~\eqref{eq_coordchangerad2harm} removes the logarithms of all the tails.
As for the other interactions, there is good hope that one can perform this simultaneous cancellation by adapting the procedure described in~\cite{Trestini:2022tot} with EM moments.
This is obviously more a wishful thinking than a formal proof, and it will have to be verified order by order.

Projecting the asymptotic expansion of the EM field~\eqref{eq_A_coordrad_infty} on the transverse basis as in Eq.~\eqref{eq_AiT}, one can easily extract the corrections to the radiative moments
\begin{subequations}\label{eq_QiHi_tail}
\begin{align}
\dQ_i & 
= \dE_i^{(1)} + \frac{2G\dM}{c^3} \int_0^\infty\!\!\!\!\!\dd\tau\bigg[\ln\left(\frac{\tau}{2b_0}\right) + \frac{3}{4}\bigg] \dE_i^{(3)}\big(U-\tau\big) + \mathcal{O}\left(\frac{G}{c^5}\right)+\calO\left(G^2\right)\,, \label{eq_Qi_tail}\\
\dH_i & 
= \dB_i^{(1)} + \frac{2G\dM}{c^3} \int_0^\infty\!\!\!\!\!\dd\tau\bigg[\ln\left(\frac{\tau}{2b_0}\right) + \frac{3}{4}\bigg] \dB_i^{(3)}\big(U-\tau\big) + \mathcal{O}\left(\frac{G}{c^5}\right) + \calO\left(G^2\right)\,. \label{eq_Hi_tail}
\end{align}
\end{subequations}
This result is only the leading order correction to the two lowest-order radiative moments entering the flux.
Note that, because of the powers in $c$ in the formulas~\eqref{eq_flux_F_EM} and~\eqref{eq_flux_Gi_EM}, the correction to $\dH_i$ enters at the same PN order than the next-to-leading corrections to $\dE_i$, namely the ``memory'' $\dE_i \times \dI_{jk}$, the ``failed-tail'' $\dE_i \times \dJ_j$ and the interactions with gauge moments, not presented here.
However, and as discussed in Sec.~\ref{subsec_Flux_toymodel}, none of those corrections enter at the precision required for our numerical applications.\\

Let us stress that this formalism is valid for any type of compact-supported and post-Newtonian sources, \emph{i.e.} enjoying weak field and small velocities.
This is a first step towards a proper description of environmental and/or physical effects.
It can thus be applied to multiple star systems embedded with EM fields, but also supernovae, ``mountainous'' neutron stars or even systems containing non-ultrarelativistic accretion disks.
In Sec.~\ref{sec_toy_model}, we apply this formalism to an astrophysically-motivated model, encompassing notably the case of binaries made of white dwarfs.

\section{Application to realistic systems}\label{sec_toy_model}

Let us now apply the wave-generation formalism constructed in the previous sections to a simple realistic model, consisting of two stars bearing magnetic dipoles.
The stars are modeled by point particles of mass $m_A$ with magnetic dipoles $\mu_A^i$ and vanishing electric dipoles $q_A^i = 0$, and we note $m = m_1 + m_2$ the total mass, $\nu = m_1m_2/m^2$ the symmetric mass ratio and $\delta = (m_1-m_2)/m$. We will use here the same setup as was done in the study of the conservative sector~\cite{Henry:2023guc}, namely two constant magnetic dipoles, aligned and normal to the orbital plane (as discussed in Sec.~IV.A of~\cite{Henry:2023guc}, the motion is indeed planar under those hypothesis) :
\begin{equation}\label{eq_muAi_ansatz}
\mu_A^i = \frac{\tilde{\mu}_A}{Gm^2}\,l^i\,,
\end{equation}
where $\boldsymbol{l}$ is the normal to the orbital plane and the normalization has been chosen for dimensional purposes.
This configuration has been shown to be the state of equilibrium for a binary system of stars with magnetic dipoles~\cite{Aykroyd:2023cvt}.
Note that, contrary to the study of the conservative sector, we can work ``on-shell'' and directly enforce~\eqref{eq_muAi_ansatz} together with $q_A^i = 0$ in the intermediate computations.

As the motion is planar, we can decompose the center-of-mass (CoM) relative velocity of the two bodies as $v^i = \dot{r}n^i + r\dot{\phi} \lambda^i$, where $\boldsymbol{\lambda}$ closes the time-dependent orthonormal basis $(\boldsymbol{n},\boldsymbol{\lambda},\boldsymbol{l})$, and we introduce the convenient shortcuts
\begin{equation}
\tilde{\mu}_+ \equiv \tilde{\mu}_1 + \tilde{\mu}_2
\qquad\text{and}\qquad
\tilde{\mu}_- \equiv \frac{m}{m_2}\, \tilde{\mu}_2 - \frac{m}{m_1}\, \tilde{\mu}_1\,.
\end{equation}
In the same spirit as~\cite{Henry:2023guc}, we seek to compute the next-to-leading (NLO) PN orders in the radiated fluxes and amplitude modes.

\subsection{Computation of the source moments}\label{subsec_source_moments}

Let us briefly comment on the computation of the source moments.\footnote{All the computations presented hereafter were performed with the use of the \textit{xAct} library from the \textit{Mathematica} software~\cite{xtensor}.}
The techniques have been widely developed in the literature and we let the interested reader refer to the references given hereafter for more details.
The first step is to parametrize the gravitational and EM fields in the near-zone in terms of so-called PN potentials. 
This parametrization was crucial for the derivation of the conservative sector, and thus has been presented in~\cite{Henry:2023guc}. 
However, in the present problem, we restrict ourselves to the computation of the NLO, and do not take the electric dipoles into account.
This implies that we require those parametrizations at a lower PN order than in the previous work~\cite{Henry:2023guc}. 
The gravitational field is thus parametrized as
\begin{subequations}\label{eq:PNmetric}
\begin{align}
h^{00} &= -\frac{4V}{c^2} - \frac{4}{c^4}\left[2V^2+\frac{W_{kk}}{2}\right]+ \frac{2\varphi^2}{c^6}+ \mathcal{O}(c^{-8})\,,\\
h^{0i} &= - \frac{4V_i}{c^3}+\mathcal{O}(c^{-5})\,,\\
h^{ij} &= -\frac{4}{c^4}\left(W_{ij}-\frac{\delta_{ij}}{2}\, W_{kk}\right)+ \mathcal{O}(c^{-6})\,,
\end{align}
\end{subequations}
and the EM field, as
\begin{equation}\label{eq:PNem}
A_0 = \frac{\varphi}{c^3}-\frac{V\varphi}{c^5} +\mathcal{O}(c^{-7})\,, \qquad A_i = \frac{\chi_i}{c^4} +\mathcal{O}(c^{-6})\,.
\end{equation}
The potentials $\{V,V_i,W_{ij},\varphi,\chi_i \}$ satisfy the wave equations presented in Eqs.~(3.7) of~\cite{Henry:2023guc}, and have been computed \emph{off-shell} to solve the conservative problem.
In the present work, we can derive them \emph{on-shell}, \emph{i.e.} we can replace the accelerations by the already derived equations of motions, and use directly the ansatz~\eqref{eq_muAi_ansatz} for $\mu_A^i$ with $q_A^i = 0$.

With this parametrization at hand, let us focus on the computation of the EM source moments. 
From their general expressions~\eqref{eq_EBK_expr}, we start by deriving their PN-expanded sources $\bar{\iota}_\mu$~\eqref{eq_iota_expl}.
To do so, we perform spatial and temporal indices decompositions and expand using the parametrization of the fields~\eqref{eq:PNmetric}--\eqref{eq:PNem}.
This procedure yields integrals with both a compact and a non-compact supported part.
The compact sectors result from the compact sources terms~\eqref{eq_Tmunu_Jmu_expl} and~\eqref{eq_Dmunu_expl}, while the non-compact sector consists of products of the potentials.
The techniques used to compute the resulting integrals are based on the \textit{Hadamard partie finie} regularization and are described in, e.g.~\cite{Blanchet:1998vx,Blanchet:2000nu,Blanchet:2004bb,Marchand:2020fpt}.
After integration, we end up with expressions of the source moments in terms of the positions and velocities of the particles, expressed in a general frame. 
The last step is thus to express them in the CoM frame, which is done by applying the method depicted in Sec. III. E. of~\cite{Henry:2023guc}.
Even in the CoM frame, the expressions of the source moments are too long to be displayed here, but they constitute only an intermediate result.
The relevant (\emph{i.e.} observable) quantities are indeed the radiative moments, which, at the NLO, are essentially their time derivatives, see Eq.~\eqref{eq_QLHL_ELBL}.

\subsection{Computation of the fluxes and modes}\label{subsec_Flux_toymodel}

Before using the method described above to compute the source moments, let us investiguate the required precision needed to reach the NLO order for magnetic effects. 
Let us first concentrate on the EM fluxes~\eqref{eq_flux_F_EM} and~\eqref{eq_flux_Gi_EM}.
A rapid inspection of the expression for the source moments~\eqref{eq_EBK_expr} in our setup reveals that the LO of the electric-type source moments $\{\dE_L\}$ start at $\calO(c^{-2})$.
Moreover, and as expected, it is easy to realize that $\dB_i = \sum_A\mu_A^i + \calO(c^{-2})$.
As $\mu_A^i$ is constant in our model, the radiative moment $\dH_i = \dB_i^{(1)} + \calO(G/c^3)$ begins at $\calO(c^{-2})$ and, due to the same fact, the associated tail~\eqref{eq_Hi_tail} is a 2.5PN relative correction.
By similar arguments, one can see that, in our setup, the EM fluxes~\eqref{eq_flux_F_EM} and~\eqref{eq_flux_Gi_EM} start at $\calO(c^{-9})$, and that non-linear interactions do not enter the NLO order, \emph{i.e.} the $\calO(c^{-11})$ coefficient in the flux.
Therefore we can approximate the radiative moments by the (time derived) source moments, as in Eq.~\eqref{eq_QLHL_ELBL}, and we only require $\{\dE_i,\dE_{ij},\dB_i,\dB_{ij},\dB_{ijk}\}$ to $\{\calO(c^{-4}),\calO(c^{-2}),\calO(c^{-2}),\calO(c^{-2}),\calO(c^{0})\}$.

As for the gravitational fluxes~\eqref{eq_flux_FGi_grav}, the magnetic effects enter in the radiative moments $\{\dU_L,\dV_L\}$ at a relative 2PN order, as was already the case in the conservative sector~\cite{Henry:2023guc}.
Therefore, the EM sector of those fluxes are also of order $\calO(c^{-9})$, and we only need the knowledge of the NLO order in $\dU_{ij}$ and the leading orders of $\dU_{ijk}$ and $\dV_{ij}$.
Similarly to the case of the EM fluxes, one can show that there is no EM corrections to the non-linear interactions up to $\calO(c^{-11})$ in the fluxes, and thus one can safely take
\begin{equation}
\dU_L = \dU_L^\text{pp} + \frac{\aem}{c^4}\,\dU_L^\text{EM} = \dI_L^{(\ell)} + \mathcal{O}(G)
\qquad\text{and}\qquad
\dV_L =  \dV_L^\text{pp} + \frac{\aem}{c^4}\,\dV_L^\text{EM} = \dJ_L^{(\ell)} + \mathcal{O}(G)\,.
\end{equation}

All the required (gravitational and EM) radiative moments were derived using the time derivatives of the source moments computed following the procedure depicted previously.
Gathering them and applying the formulas~\eqref{eq_flux_F_grav} and~\eqref{eq_flux_F_EM}, the total energy flux\footnote{The Noetherian fluxes given in Eqs.~(3.14) of~\cite{Henry:2023guc} are vanishing in our configuration.} is given by
\begin{equation}
\mathcal{F} = \mathcal{F}_\text{GW} + \mathcal{F}_\text{EM} = \mathcal{F}_\text{GW,pp} + \mathcal{F}_\text{GW,mag} + \mathcal{F}_\text{EM}\,,
\end{equation}
where our nomenclature should be transparent.
Note that the point-particle contribution $\mathcal{F}_\text{GW,pp}$ is currently known at a relative 4PN order~\cite{Blanchet:2023sbv}, but is only required at NLO in this work.
Similarly, working out the formulas~\eqref{eq_flux_Gi_grav} and~\eqref{eq_flux_Gi_EM}, we find that the total angular momentum flux is directed along $\boldsymbol{l}$, and read
\begin{equation}
\mathcal{G}^i = \mathcal{G}^i_\text{GW} + \mathcal{G}^i_\text{EM} = \bigg[\mathcal{G}_\text{GW,pp} + \mathcal{G}_\text{GW,mag} + \mathcal{G}_\text{EM}\bigg]l^i\,.
\end{equation}
The explicit expressions of each contribution to NLO are displayed in App.~\ref{app_expr_flux}.
Then, using the formulas~\eqref{eq_mode_hlm2ULVL} and~\eqref{eq_alphalmL_def}, one can extract the EM corrections to the gravitational modes, presented in App.~\ref{app_modes}.
Those results are collected in the ancillary file~\cite{AncFile}.

\subsection{The case of eccentric orbits}\label{subsec_elliptic}

All the results presented in the previous section~\ref{subsec_Flux_toymodel} are given for generic (planar) orbits in terms of $(r,\dot{r},\dot{\phi})$.
We will now specify them on quasi-elliptic trajectories in order to extract physical information on the evolution of the orbital parameters.

\subsubsection{Quasi-Keplerian parametrization of the orbits}\label{subsubsec_QK_par}

In~\cite{Henry:2023guc}, only quasi-circular orbits were studied.
Thus, we first need to derive the quasi-Keplerian parametrization of the trajectories at the required order before proceeding any further.
In a post-Newtonian framework, celestial bodies are naturally expected to follow nearly Keplerian orbits.
For spinless point particles, the first corrections to the Keplerian orbits were computed in~\cite{DamourpK}, then extended at 2PN~\cite{Schaefer:1993}, 3PN~\cite{Memmesheimer:2004cv}, and finally up to the current 4PN precision~\cite{Cho:2021oai}.
The aim of this section is to derive a quasi-Keplerian representation of the motion at NLO, by taking the EM corrections into account.

Following the standard procedure, depicted \emph{e.g.} in~\cite{Tessmer:2010hp,Henry:2023tka} (notably, all required integrals are displayed in App.~A\footnote{Note a missing $e_r$ in front of the $\sin u$ of (A.7) of~\cite{Tessmer:2010hp}.} of~\cite{Tessmer:2010hp}), we can establish the relations between the orbital separation $r$, the phase $\phi$, the mean anomaly $\ell$, the eccentric anomaly $u$ and the true anomaly $v$.
They have the same structure as the pure GR ones at 2PN, namely
\begin{subequations}\label{eq_qK_parametrisation}
\begin{align}
&
r = a_r \big(1- e_r\,\cos u \big)\,,\\
&
\ell \equiv n\big(t-t_0\big) = u - e_t\,\sin u + f_{v-u}\big(v-u\big) + f_v\,\sin v\,,
\label{eq_qK_parametrisation_ell}\\
&
\frac{\phi-\phi_0}{K} = v +g_{2v}\,\sin(2v) + g_{3v}\,\sin(3v)\,,\\
&
v = 2 \arctan\bigg[\sqrt{\frac{1+e_\phi}{1-e_\phi}}\,\tan\frac{u}{2}\bigg]\,,
\end{align}
\end{subequations}
where $n = \frac{2\pi}{P}$, with $P$ the relativistic period, and $K = 1 + k$, with $k$ the advance of periastron.
The expressions of all parameters entering Eq.~\eqref{eq_qK_parametrisation} are given in terms of the invariant energy and angular momentum in App.~\ref{app_qK_par}, and collected in the ancillary file~\cite{AncFile}.
Note that the coefficients translating a departure from a Kepler-like structure, $f_\star$ and $g_\star$ Eqs.~\eqref{eq_qK_Fvu}--\eqref{eq_qK_G3v}, are all starting at the same order, namely $\mathcal{O}(\aem/c^6)$.
The relations~\eqref{eq_qK_parametrisation} depict entirely the conservative motion of the binary on a quasi-elliptic orbit, and as such, contains all the information we need in the following section.

\subsubsection{Evolution of the orbital parameters}\label{subsec_QK_evol}

Instead of using the gauge-invariant energy and angular momentum, we will describe the evolution of the orbital parameters in terms of the time-eccentricity $e_t$ and the orbital frequency-related parameter
\begin{equation}\label{eq_xdef_qK}
x = \left(\frac{Gm\Omega}{c^3}\right)^{2/3}\,,
\end{equation}
where $\Omega = Kn = \frac{2\pi}{P}(1+k)$. 
From Eqs.~\eqref{eq_qK_parametrisation}, one can derive $r(u)$, $\dot{r}(u)$ and $\dot{\phi}(u)$ using chain rules. We finally obtain their expressions in terms of $(x,e_t,u)$, which allows to plug them into the fluxes~\eqref{eq_energy_flux_gen} and~\eqref{eq_ang_flux_gen} to get the exact instantaneous fluxes $\mathcal{F}$ and $\mathcal{G}$ in terms of the eccentric anomaly.
With such expressions at hand, one can perform their orbit average as
\begin{equation}
\left\langle \mathcal{F}\right\rangle \equiv \frac{1}{P}\int_0^P\!\!\dd t\,\mathcal{F}(t) = \frac{1}{2\pi}\int_0^{2\pi}\!\!\dd u\,\frac{\dd \ell}{\dd u}\,\mathcal{F}(u)\,,
\end{equation}
and similarly for $\left\langle \mathcal{G}\right\rangle$.
Using the integration formula (8.5) of~\cite{Arun:2007sg}
\begin{equation}
\frac{1}{2\pi}\int_0^{2\pi}\!\!\frac{\dd u}{(1-e_t\,\cos u)^n} = \frac{1}{(1-e_t^2)^{n/2}}\,P_{n-1}\bigg(\frac{1}{\sqrt{1-e_t^2}}\bigg)\,,
\end{equation}
with $P_n$ the usual Legendre polynomial, we derived the expressions for the orbital averaged fluxes, presented in App.~\ref{app_qK_flux}.
Finally, one can use the fluxes together with the explicit expressions~\eqref{eq_qK_ar}, \eqref{eq_qK_n}, \eqref{eq_qK_K} and~\eqref{eq_qK_et} to derive the averaged evolution of the orbital parameters $e_r$, $a_r$ and $x$.
Indeed, the flux-balance equations~\eqref{eq_balance_eqs} allow us to derive the chain rule for, \emph{e.g.}, the time eccentricity
\begin{equation}\label{eq_dedt_chain_rule}
\left\langle\frac{\dd e_t}{\dd t}\right\rangle = -\frac{\partial e_t}{\partial E}\left\langle \mathcal{F}\right\rangle-\frac{\partial e_t}{\partial J}\left\langle \mathcal{G}\right\rangle\,,
\end{equation}
where we have denoted the norm of the conserved angular momentum as $J = \vert J^i \vert$.
Those generalization of the Peters \& Mathews formulas~\cite{Peters:1963ux,Peters:1964zz} can be found in App.~\ref{app_qK_orbital} as well as in the ancillary file~\cite{AncFile}.

\subsubsection{Gravitational modes}\label{subsec_QK_mode}

In order to obtain a full waveform, we have also computed the gravitational amplitude modes for quasi-elliptic orbits in terms of the eccentric anomaly $u$, using the same techniques as those employed to derive the instantaneous expressions of the fluxes.
The benefit of such parametrization in place of an expression in terms of the time (or equivalently the mean anomaly $\ell$) is that it is exact.
In order to express them in terms of $\ell$, one needs to invert the relation $\ell(u)$~\eqref{eq_qK_parametrisation_ell}, which involves an expansion in the time eccentricity $e_t$, see \emph{e.g.}~\cite{BrouwerClemence}.

The expressions of the instantaneous part of the amplitude modes are too lengthy to be displayed, even in an Appendix, but can be found in the ancillary file~\cite{AncFile}, and we find full agreement between our point-particle 1PN sector and the results of~\cite{Henry:2023tka}.

\subsection{Quasi-circular orbits : gravitational phase and modes}\label{subsec_circ}

Finally, let us study the system on quasi-circular orbits. For most of the relevant quantities, one would only need to set the eccentricity to vanish to obtain the results for quasi-circular orbits. However, the explicit expression of the phase evolution of the system in terms of the orbital frequency is not provided within the quasi-Keplerian framework. To express the phase, we take $\dot{r}=0$ as it is a higher 2.5PN correction, and set $\dot{\phi}=\omega$ where $\omega$ is the orbital frequency for a quasi-circular motion. In such a configuration, the fluxes, amplitudes and phase only depend on the gauge invariant PN parameter
\begin{equation}
x = \left(\frac{Gm\omega}{c^3}\right)^{2/3}\,.
\end{equation}
This variable is naturally the circular limit of the one defined in the eccentric case Eq.~\eqref{eq_xdef_qK}.
The two flux-balance equations~\eqref{eq_balance_eqs} are now degenerate and the fluxes are indirectly linked by the  ``thermodynamic'' relation
\begin{equation}
\frac{\dd E}{\dd \omega} = \omega\,\frac{\dd J}{\dd \omega}\,.
\end{equation}
This relation is a consequence of the first law of binary black hole mechanics~\cite{Friedman:2001pf,LeTiec:2011ab} for systems with constant individual masses, and applies for the conserved quantities of the system. Making use of the flux-balance equations, we checked that it holds for the radiative sector. Thus we can focus solely on the flux-balance equation for the energy to derive the phase.
Using the generalized Kepler's law, Eq.~(4.5) of~\cite{Henry:2023guc}, on can express the flux in terms of $x$ at NLO
\begin{equation}
\begin{aligned}
\mathcal{F} = \frac{32c^5\nu^2x^5}{5G}\Bigg\lbrace
&
1 -\bigg(\frac{1247}{336}+\frac{35\nu}{12}\bigg)x 
-4\aem \Bigg[\tilde{\mu}_+^2 + \delta\,\tilde{\mu}_+\tilde{\mu}_--\bigg(\frac{1}{96}+\nu\bigg)\tilde{\mu}_-^2\Bigg]x^2\\
&
- \aem \Bigg[
\bigg(\frac{755}{168}-\frac{169\nu}{6}\bigg)\tilde{\mu}_+^2
+ \bigg(\frac{685}{168}-\frac{169\nu}{6}\bigg)\delta\,\tilde{\mu}_-\tilde{\mu}_+\\
&
\qquad \qquad-\bigg(\frac{19}{32}+\frac{3193\nu}{1008}- \frac{169\nu^2}{6}\bigg)\tilde{\mu}_-^2 \Bigg]x^3\Bigg\rbrace\,.
\end{aligned}
\end{equation}
We naturally recover~\eqref{eq_qK_F_avg} for $e_t=0$. From this quantity and the expression of the energy in terms of $x$, Eq.~(4.6a) of~\cite{Henry:2023guc}, one can extract the gravitational phase as
\begin{equation}\label{eq_phi_def}
\phi = \int\!\!\dd t\,\omega = - \frac{c^3}{Gm}\int\!\!\dd x\,\frac{x^{3/2}}{\mathcal{F}}\,\frac{\dd E}{\dd x}\,.
\end{equation}
At NLO, it comes
\begin{equation}\label{eq_phase_circ}
\begin{aligned}
\phi = \phi_0 - \frac{1}{32\nu x^{5/2}}
\Bigg\lbrace
&
1 + \bigg(\frac{3715}{1008}+\frac{55\nu}{12}\bigg)x
+50\,\aem \Bigg[\tilde{\mu}_+^2+\delta\,\tilde{\mu}_+\tilde{\mu}_--\bigg(\frac{1}{240}+\nu\bigg)\tilde{\mu}_-^2\Bigg]x^2\\
&
- \aem \Bigg[
\bigg(\frac{5865}{14}+\frac{280\nu}{3}\bigg)\tilde{\mu}_+^2
+ \bigg(\frac{35015}{84}+\frac{280\nu}{3}\bigg)\delta\,\tilde{\mu}_+\tilde{\mu}_-\\
&\qquad\qquad
- \bigg(\frac{16945}{4032}+\frac{416785\nu}{1008}+ \frac{280\nu^2}{3}\bigg)\tilde{\mu}_-^2\Bigg]x^3\Bigg\rbrace\,.
\end{aligned} 
\end{equation}
where $\phi_0$ is a constant of integration and the phase has been collected in the ancillary file~\cite{AncFile}.

Then, we can also express the gravitational modes in circular orbits. 
For the sake of simplicity, we rescale the observable gravitational modes $h_{\ell \dm}$~\eqref{eq_mode_hlm2ULVL} by the dominant contribution, and single out the dependence on the gravitational phase $\phi$ as
\begin{equation}\label{eq_mode_Hlm_circ}
h_{\ell \dm}
= \frac{8 G m \nu x}{c^2 R} \,\sqrt{\frac{\pi}{5}} \,H_{\ell \dm} \, \de^{-\di \dm \phi}\,.
\end{equation}
Note that the coefficients $H_{\ell \dm}$ for circular orbits differ from the ones defined in generic orbits, $\mathcal{H}_{\ell\dm}$~\eqref{eq_mode_Hlm_gen} by a factor $c^2 x$.
The relation $H_{\ell,-\dm} = (-)^\ell \,\overline{H}_{\ell \dm}$, where the overbar denotes the complex conjugate, still holds, thus we only present the coefficients with positive $\dm$.
At the relevant order, the instantaneous modes read
\begin{subequations}
\begin{align}
&
H_{22} =
1 - \bigg(\frac{107}{42}- \frac{55\nu}{42}\bigg) x 
- \frac{2\aem \tilde{\mu}_1\tilde{\mu}_2}{\nu}x^2\bigg[1 - \bigg(\frac{1}{21}-\frac{73\nu}{42}\bigg)x\bigg]\,,\\
&
H_{21} =
\frac{\di}{3}\,\delta\,x^{1/2}\bigg[1
- 3\,\frac{\aem \tilde{\mu}_1\tilde{\mu}_2}{\nu}\,x^2\bigg]\,,\\
&
H_{33} =
- \frac{3\di}{4}\sqrt{\frac{15}{14}}\,\delta\,x^{1/2}\bigg[1- 3\,\frac{\aem \tilde{\mu}_1\tilde{\mu}_2}{\nu}\,x^2\bigg]\,,\\
&
H_{32} =
\frac{1}{3}\sqrt{\frac{5}{7}}\big(1-3\nu\big)\,x\,\bigg[1
- 4\,\frac{\aem \tilde{\mu}_1\tilde{\mu}_2}{\nu}\,x^2\bigg]
\,,\\
&
H_{31} =
\frac{\di}{12\sqrt{14}}\,\delta\,x^{1/2}\bigg[1- 3\,\frac{\aem \tilde{\mu}_1\tilde{\mu}_2}{\nu}\,x^2\bigg]\,,\\
&
H_{44} = 
-\frac{8}{9}\sqrt{\frac{5}{7}}\big(1-3\nu\big)\,x\,\bigg[1- 4\,\frac{\aem \tilde{\mu}_1\tilde{\mu}_2}{\nu}\,x^2\bigg]\,,\\
&
H_{42} = 
\frac{\sqrt{5}}{63}\big(1-3\nu\big)\,x\,\bigg[1- 4\,\frac{\aem \tilde{\mu}_1\tilde{\mu}_2}{\nu}\,x^2\bigg]\,,
\end{align}
\end{subequations}
and all $H_{\ell,0}=0$.\footnote{This is due to the fact that, in this paper, we only consider instantaneous contributions to the modes. In fact, there are contributions of the memory to the $(\ell,0)$ modes for even $\ell$ starting at Newtonian order. In the case of eccentric orbits, these memory contributions also contribute to oscillating modes and could contribute to the magnetic sector of $h_{22}$ at NLO. We neglect these contributions throughout this paper.}

\subsection{Numerical applications}\label{subsec_numeric}

One of the motivations for this work is to estimate the significance of the magnetic contribution to the dissipative sector.
For instance, it has been shown in~\cite{Savalle:2023zbb} that EM effects on eccentric orbits were not negligible for LISA data analysis, when accumulating a sufficient observation time.
Indeed, eccentricity affects each harmonics differently, breaking the degeneracy and making the presence of magnetic dipoles possibly detectable.
However, the analysis done in~\cite{Savalle:2023zbb} relies solely on the leading order of the Peters \& Mathews formulas for point-particles~\cite{Peters:1963ux,Peters:1964zz}, neglecting EM effects in the dissipative sector.
This approach is not formally consistent in a PN sense.
In the present work, we have generalized the Peters \& Mathews formulas by including the magnetic dipoles to NLO, see Eqs.~\eqref{eq_detdt_split} and~\eqref{eq_dardt_split}, which completes the model used in~\cite{Savalle:2023zbb}. 

We can estimate the order of magnitude of the newly computed EM terms for realistic systems, by comparing them to the pure point-particle contribution.
To this end, we choose an astrophysically motivated model, that relates the magnitude of the magnetic dipole of a star to its magnetic field at its surface and its radius as~\cite{Pablo_2019}
\begin{equation}\label{eq_mag_model_astr}
\vert \mu_A^i \vert = \frac{2\pi}{\mu_0}\,B_A R_A^3\,.
\end{equation}
As for the numerical values of the masses and magnetic fields of the stars, we took those already used in~\cite{Bourgoin:2022ilm,Henry:2023guc}, which are typical values for white dwarfs, see Table~\ref{tab_num_param}.
\begin{table}[!h]
    \centering
    \begin{tabular}{|c|c|c|c|}
    \hline
   Parameter & Unit &  Value for the first star & Value for the second star \\
    \hline
    \hline
    $m_A$ & $M_\odot$ & $1.2$ & $0.3$ \\
    \hline
    $R_A$ & km & $6.0\cdot 10^3$ & $15 \cdot 10^3$ \\
    \hline
    $B_A$ & G & $1.0\cdot10^9$ & $1.0\cdot10^9$ \\
    \hline
    \end{tabular}
    \caption{Numerical values taken for the physical parameters of each star, reproduced from~\cite{Bourgoin:2022ilm}. $m_A$ is the mass of the star $A$, $R_A$ its equatorial radius, and $B_A$ the magnitude of the magnetic field at its surface.}
    \label{tab_num_param}
\end{table}

Finally, we selected three different configurations for the fundamental observed frequency (which is half the orbital frequency), $10^{-1}$, $10^{-2}$ and $10^{-3}$ Hz. 
Those values are motivated by the bandwidth of the LISA detector~\cite{LISA:2017pwj}.
Numerical estimates of the magnitude of the magnetic contributions to $\langle\dot{a}_r\rangle$ and $\langle\dot{e}_t\rangle$ are provided in Table~\ref{tab_adot_edot}, and it is clear that these corrections, of order at most $10^{-6}$, are quite small, even in the extreme case $e_t = 0.5$.
This suggests that the model used in~\cite{Savalle:2023zbb} would yield qualitatively similar conclusions if the correct evolution equations (taking into account the magnetic corrections) were considered.
\begin{table}[!h]
    \centering
    \begin{tabular}{|c||c|c|c||c|c|c|}
    \hline
    & \multicolumn{3}{ c|| }{$\vert\langle\dot{a}_r\rangle^\text{EM}/\langle\dot{a}_r\rangle^\text{GR}\vert$} &\multicolumn{3}{ c| }{$\vert\langle\dot{e}_t\rangle^\text{EM}/\langle\dot{e}_t\rangle^\text{GR}\vert$} \\
    \hline
    $e_t$ & Config. 1 & Config. 2 & Config. 3 & Config. 1 & Config. 2 & Config. 3 \\
    \hline
    \hline
    0.1 & $7.0\cdot 10^{-7}$ & $3.3\cdot 10^{-8}$ & $1.5\cdot 10^{-9}$ & $6.8\cdot 10^{-7}$ & $3.2\cdot 10^{-8}$ & $1.5\cdot 10^{-9}$ \\
    \hline
    0.3 & $9.8\cdot 10^{-7}$ & $4.6\cdot 10^{-8}$ & $2.1\cdot 10^{-9}$ & $8.9\cdot 10^{-7}$ & $4.1\cdot 10^{-8}$ & $1.9\cdot 10^{-9}$ \\
    \hline 
    0.5 & $1.8\cdot 10^{-6}$ & $8.2\cdot 10^{-8}$ & $3.8\cdot 10^{-9}$ & $1.5\cdot 10^{-6}$ & $7.1\cdot 10^{-8}$ & $3.3\cdot 10^{-9}$ \\
    \hline
    \end{tabular}
    \caption{Numerical estimation of the ratios $\vert\langle\dot{a}_r\rangle^\text{EM}/\langle\dot{a}_r\rangle^\text{GR}\vert$ and $\vert\langle\dot{e}_t\rangle^\text{EM}/\langle\dot{e}_t\rangle^\text{GR}\vert$ for different values of the orbital frequency and eccentricities. Configuration 1 corresponds to $\Omega = 5\cdot 10^{-2}$Hz, configuration 2, to  $\Omega = 5\cdot 10^{-3}$Hz and configuration 3, to  $\Omega = 5\cdot 10^{-4}$Hz, all of those based on the values of Table~\ref{tab_num_param}.}
    \label{tab_adot_edot}
\end{table}

As was the case in the conservative sector~\cite{Henry:2023guc}, the leading EM corrections are smaller than the NLO point-particle ones, but larger than the NNLO point-particle ones, and this for each of the different chosen configurations. 
This makes the EM contributions roughly comparable to a 1.5PN point-particle term.
On the other hand, the EM contributions tend to dominate over the point-particle ones for eccentricities close to 1.
But, if our formulas~\eqref{eq_detdt_split} and~\eqref{eq_dardt_split} are formally exact for any bound orbit (\emph{i.e.} $0\leq e_t<1$) within the PN regime, it is a well-known fact that the PN approximation breaks down for high eccentricities.
This is due to the fact that, at the periastron of a highly eccentric orbit, the weak-field condition is naturally invalid.
Thus, the enhancement of magnetic effects for high eccentricities is spurious.
Furthermore, when the two stars are close enough, depicting the magnetic interaction with dipoles is no more possible, as other effects (such mass exchange or common magnetic envelop) enter the picture.

Finally, let us comment on the application to binaries of neutron stars.
As discussed in Sec.~\ref{sec_introduction}, those have much stronger magnetic fields than white dwarfs.
However, their magnetic dipoles are significantly smaller, due to the fact that their radius are $\sim10^{3}$ shorter than those of white dwarfs, see Eq.~\eqref{eq_mag_model_astr}.
Thus, in our setup of constant and aligned magnetic dipoles, the EM contributions to the decay of orbital parameters for binaries of neutron stars systems are expected to be  negligible.

\section{Summary and conclusion}\label{sec_CCL}

Including electromagnetic effects in the templates used by future gravitational-wave detectors, such as LISA or ET, will be important, notably for an accurate characterization of white dwarfs binaries~\cite{Bourgoin:2021yzf,Bourgoin:2022ilm,Bourgoin:2022qex,Bourgoin:2022ibr,Carvalho:2022pst,Savalle:2023zbb,Aykroyd:2023cvt}.
In this spirit, a proper post-Newtonian treatment of the motion of two stars with EM dipoles has been done in~\cite{Henry:2023guc}, and the usual Noetherian quantities (energy, angular momentum, \emph{etc.}) have been derived.
Pushing further in this direction, the present work tackles the problem of the generation of gravitational-wave by such binaries.

Relying on matched multipolar-post-Minkowskian and post-Newtonian expansions, we first concentrate on the near-zone behavior of the fields, and derive \emph{source} and \emph{gauge} moments~\eqref{eq_EBK_expr}, that entirely describe the physics of the matter distribution.
From those moments, we iteratively construct non-linear \emph{radiative} moments that parametrize the values of the fields at future null infinity,~\eqref{eq_HijTT} and~\eqref{eq_AiT}.
This asymptotic parametrization allows us to extract the fluxes of energy and angular momentum~\eqref{eq_flux_FGi_grav}, \eqref{eq_flux_F_EM} and~\eqref{eq_flux_Gi_EM}, as well as the gravitational modes, parametrizing the waveform as detected by an observer at infinity.
As a proof of concept, we apply the iteration scheme to explicitly derive the EM \emph{tail}, \emph{i.e.} the leading order non-linear correction, of the two lowest-order EM radiative moments~\eqref{eq_QiHi_tail}. 
We emphasize that this formalism is valid for any type of post-Newtonian, compact-supported source (\emph{i.e.} enjoying weak field and small velocities).
It can thus be applied to other astrophysical systems containing EM effects within the PN regime, for instance non-spherical rotating neutron stars, supernovae or even systems containing non-ultrarelativistic accretion disks.

Next, we apply this formalism to the concrete case of a binary composed of stars bearing magnetic dipoles.
Under the assumption of constant and aligned magnetic dipoles, we derive the fluxes and modes for generic orbits.
Enforcing eccentric orbits, we derive the evolution of orbital parameters (eccentricity, semi-major axis and frequency)~\eqref{eq_detdt_split}, \eqref{eq_dardt_split} and~\eqref{eq_dxdt_split}, and provide the EM corrections to the gravitational modes to the next-to-leading order.
For the specific case of quasi-circular orbits, we derive the correction to the gravitational phase.
All those results are collected in the ancillary file~\cite{AncFile}.

Specifying realistic astrophysical configurations, we perform numerical estimates that show the magnitude of the magnetic contribution to the radiated gravitational waves, and dissipative dynamics. 
Notably, we find that the ratio of the magnetic effects \emph{vs.} the pure point-particle terms in the secular evolution of the orbital elements are at most $\sim 10^{-6}$. This implies that the model used in the study~\cite{Savalle:2023zbb} would yield the qualitatively similar conclusions if it were to be consistent in a post-Newtonian sense. 
However, and similarly to~\cite{Henry:2023guc}, we find that the leading magnetic corrections are smaller than a 1PN point-particle effect, but bigger than a 2PN one, making the magnetic effects roughly comparable to a 1.5PN quantity.

If the present work formally tackles the problem of post-Newtonian gravitational wave generation in an Einstein-Maxwell framework, the efforts in building realistic waveform models including EM effects can naturally be improved.
In fact, our model assumes constant and aligned magnetic dipoles.
If such configuration is indeed an equilibrium state~\cite{Aykroyd:2023cvt}, it would be quite interesting to apply our formalism to systems with non-constant and/or non-aligned dipoles.
In such configurations, we could have quantitative changes, as the leading order in the EM fluxes~\eqref{eq_flux_F_EM} and~\eqref{eq_flux_Gi_EM} is vanishing due to the constant dipole hypothesis.
This requires to fix the dynamics of the dipoles, which could be done for instance by coupling them to the spins of the stars. This is left for future work.

\acknowledgments

It is a pleasure to thank C. Dlapa, G. Faye, M. Riva and L. Speri for interesting discussions.
F.L. received funding from the European Research Council (ERC) under the European Union’s Horizon 2020 research and innovation program (grant agreement No 817791). 
C.LPL. acknowledges the financial support of Centre National d'\'{E}tudes Spatiales (CNES) for LISA.

\appendix

\section{Electro-magnetic fluxes}\label{app_EM_fluxes}

This appendix discusses the construction of the EM fluxes~\eqref{eq_flux_F_EM} and~\eqref{eq_flux_Gi_EM} in terms of the radiative moments.

\subsection{Generic expressions}\label{app_EM_fluxes_gen}

Flux balance equations within the PN-MPM formalism have been treated in, \textit{e.g.} ~\cite{Blanchet:2018yqa}. In this reference, the radiated fluxes are defined using the  angular integral of the Landau-Lifshitz pseudotensor in the radiative zone. In the present paper, this tensor is now composed of the pure GR and the EM parts. However, as discussed in Sec.~\ref{subsec_iteration}, both fields bear the same asymptotic multipolar structure. Thus, the same defintion holds when generalizing to EM effects. This allows us to directly use Eqs. (4.4) of~\cite{Blanchet:2018yqa} by defining the total energy and angular momentum fluxes as
\begin{subequations}
\begin{align}
&
\mathcal{F} \equiv \lim_{R\to\infty} c\,R^2 \int\!\!\dd^2\Omega\,N_i\tau^{0i}\,,\\
&
\mathcal{G}^i \equiv \lim_{R\to\infty} R^3\,\varepsilon_{iab} \int\!\!\dd^2\Omega\,N_{ac}\tau^{bc}\,.
\end{align}
\end{subequations}
The compact sector of $\tau^{\mu\nu}$~\eqref{eq_tau_expl} cannot contribute to the surface integrals, so we are left with the contribution of $\Lambda^{\mu\nu} =  \Lambda_g^{\mu\nu} + \Lambda_\text{EM}^{\mu\nu}$.
The contribution of $\Lambda_g^{\mu\nu}$ has been treated in~\cite{Thorne:1980ru} and gives the usual gravitational fluxes~\eqref{eq_flux_FGi_grav}.
Focusing on $\Lambda_\text{EM}^{\mu\nu}$, we define the ``EM'' part of the fluxes as
\begin{subequations}
\begin{align}
\label{eq_app_flux_FEM}
&
\mathcal{F}_\text{EM} \equiv \lim_{R\to\infty}\left[ \frac{c^5\,R^2}{16\pi G} \int\!\!\dd^2\Omega\,N_i\Lambda_\text{EM}^{0i} \right]
\underrel{R\to\infty}{=}\frac{c^4\,R^2}{4\pi G \aem} \int\!\!\dd^2\Omega\,N_i\,\Pi_i\,,\\
&
\label{eq_app_flux_GiEM}
\mathcal{G}_\text{EM}^i 
\equiv \lim_{R\to\infty}\left[\frac{c^4\,R^3}{16\pi G}\,\varepsilon_{iab}\int\!\!\dd^2\Omega\,N_{ac}\,\Lambda_\text{EM}^{bc}\right]
\underrel{R\to\infty}{=} \frac{c^4\,R^3}{4\pi G\aem}\,\varepsilon_{iab}\int\!\!\dd^2\Omega\,N_{ac}\,\bigg(F_{bd}F_{cd} - F_{0b}F_{0c}\bigg)\,,
\end{align}
\end{subequations}
where we recall $\Pi_i = c\,F_{ij}F_{j0}$ is the Poynting vector.

\subsection{Gauge-independent derivation of the energy flux}\label{app_EM_fluxes_F}

In Sec.~\ref{subsubsec_flux_EM}, the energy flux has been derived from the asymptotic expansion of $A_\mu$ in a transverse gauge.
In order to emphasize that our result~\eqref{eq_flux_F_EM} is indeed gauge-independent, let us re-derive it using a generic (\emph{i.e.} not transverse) form of the EM field
\begin{subequations}\label{eq_app_A_asympt_NT}
\begin{align}
&
A_0 = - \frac{G\aem}{c^3\,R} \sum_{\ell\geq1} \frac{1}{c^\ell\,\ell !}N_L \dQ_L+ \mathcal{O}\left(\frac{1}{R^2}\right)\,,\\
&
A_i = \frac{G\aem}{c^3\,R} \sum_{\ell\geq1} \frac{1}{c^\ell\,\ell !}\bigg[N_{L-1} \dQ_{iL-1} - \frac{\ell}{c(\ell+1)}\,\varepsilon_{iab}N_{aL-1}\dH_{bL-1}\bigg]+ \mathcal{O}\left(\frac{1}{R^2}\right)\,,
\end{align}
\end{subequations}
where the moments are naturally evaluated in the (radiative) retarded time $T-R/c$.
After some massaging, using notably the relations
\begin{subequations}
\begin{align}
&
\partial_i \Bigg(\frac{N_L}{R}\,\dQ_L\Bigg) = 
- \frac{N_{iL}}{R\,c}\,\dot{\dQ}_L  + \mathcal{O}\left(\frac{1}{R^2}\right)\,,\\
&
\partial_i \Bigg(\varepsilon_{jab}\,\frac{N_{aL-1}}{R}\,\dH_{bL-1}\Bigg) = - \varepsilon_{jab}\,\frac{N_{iaL-1}}{R\,c}\,\dot{\dH}_{bL-1} + \mathcal{O}\left(\frac{1}{R^2}\right)\,,
\end{align}
\end{subequations}
the field strength reads at leading order
\begin{subequations}
\begin{align}
&
F_{0i} 
= \frac{G\aem}{c^4\,R} \sum_{\ell\geq1} \frac{1}{c^\ell\,\ell !}\bigg[\perp_{ij}N_{L-1} \dot{\dQ}_{jL-1} - \frac{\ell}{c(\ell+1)}\,\varepsilon_{iab}N_{aL-1}\dot{\dH}_{bL-1}\bigg]\,,\\
&
F_{ij}
= \frac{G\aem}{c^4\,R} \sum_{\ell\geq1} \frac{1}{c^\ell\,\ell !}\bigg[N_{jL-1} \dot{\dQ}_{iL-1} -N_{iL-1} \dot{\dQ}_{jL-1} - \frac{\ell\,N_{aL-1}}{c(\ell+1)}\,\big(\varepsilon_{iab}N_j-\varepsilon_{jab}N_i\big)\dot{\dH}_{bL-1}\bigg]\,,
\end{align}
\end{subequations}
where we recall that the transverse projector is $\perp_{ij} = \delta_{ij} - N_{ij}$.
At the leading order, the Poynting vector $\Pi_i = c\,F_{ij}F_{j0}$ becomes
\begin{equation}\label{eq_app_Poynting}
\Pi_i = \frac{G^2\aem^2}{c^7\,R^2}\sum_{\ell,k}\frac{N_iN_{L-1}N_{K-1}}{c^{\ell+k}\,\ell!\,k!} \,\perp_{ab}\,\Bigg[\dot{\dQ}_{aL-1}\dot{\dQ}_{bK-1} + \frac{\ell k}{c^2(\ell+1)(k+1)}\,\dot{\dH}_{aL-1}\dot{\dH}_{bK-1}\bigg]\,,
\end{equation}
where we have not written the cross terms $\propto \dQ_L\dH_K$, as their odd parity will eliminate them when performing the angular integrals.
The energy flux~\eqref{eq_app_flux_FEM} then reads
\begin{equation}
\begin{aligned}
\mathcal{F}_\text{EM} 
&
=
\frac{G\aem}{4\pi c^3}\sum_{\ell,k}\frac{1}{c^{\ell+k}\,\ell!\,k!}\,\Bigg[\dot{\dQ}_{aL-1}\dot{\dQ}_{bK-1} + \frac{\ell k}{c^2(\ell+1)(k+1)}\,\dot{\dH}_{aL-1}\dot{\dH}_{bK-1}\bigg]\int\!\!\dd^2\Omega\perp_{ab}N_{L-1}N_{K-1}\\
&
= 
\sum_{\ell\geq 1} \frac{G \aem}{c^{3+2\ell}}\frac{\ell+1}{\ell\,\ell !(2\ell+1)!!}\bigg[\dot{\dQ}_L\dot{\dQ}_L + \frac{\ell^2}{c^2(\ell+1)^2}\,\dot{\dH}_L\dot{\dH}_L\bigg]
\end{aligned}
\end{equation}
where we have used the fact that the moments $\{\dQ_L, \dH_L\}$ are STF and the integration formula 
\begin{equation}\label{eq_app_angular_perp}
\dot{\dQ}_{aL-1}\dot{\dQ}_{bK-1}\int\!\!\dd^2 \Omega\,\perp_{ab} N_{L-1}N_{K-1} =  \frac{4\pi\,(\ell+1) !}{\ell\,(2\ell+1)!!}\,\delta_{k\ell}\,\dot{\dQ}_L\dot{\dQ}_L\,.
\end{equation}
This generic-gauge derivation agrees with the result~\eqref{eq_flux_F_EM}, performed in the transverse gauge, which was expected as the energy flux should be a gauge-invariant quantity.

\subsection{Angular momentum flux}\label{app_EM_fluxes_G}

Let us now turn to the more subtle derivation of the angular momentum flux.
As clear from the $R^3$ appearing in its expression~\eqref{eq_app_flux_GiEM}, and as discussed in lengths in \emph{e.g.}~\cite{Thorne:1980ru,Bonga:2018gzr,Blanchet:2018yqa}, its computation requires the knowledge of the sub-leading terms $\mathcal{O}(1/R^2)$ of the radiative field~\eqref{eq_A_coordrad_infty}.
The canonical way of proceeding involves a full analysis of the asymptotic structure at future null infinity, and is plagued by subtleties, see \emph{e.g.}~\cite{Bonga:2018gzr}.
Instead, we adopt a more pragmatic approach, following the spirit of~\cite{Blanchet:2018yqa}.
We start by computing the ``source'' flux, \emph{i.e.} the dominant PM order of the flux, by using the linearized metric~\eqref{eq_Amu_funcEBK}.
This partial flux will be written in terms of (derivatives) of the source moments $\dE_L$ and $\dB_L$, hence the name.
In order to reconstruct the ``full'' flux from this source flux, one should in principle add all higher PM orders \emph{via} tedious computations.
Nevertheless, using the argument of~\cite{Blanchet:2018yqa}, we claim that this source flux has the same structure that the ``radiative'' flux, under the naive replacements $\big(\dE_L^{(\ell)},\dB_L^{(\ell)}) \to (\dQ_L,\dH_L)$, see Eq.~\eqref{eq_QLHL_ELBL}.
This method, although not a formal proof, has the advantage of the simplicity: the linear metric~\eqref{eq_Amu_funcEBK} is easily expanded at sub-leading order, using
\begin{equation}
\hat{\partial}_L\bigg[\frac{F(t-r/c)}{r}\bigg] = \frac{(-)^\ell\hat{n}_L}{r\,c^\ell}\bigg[F^{(\ell)}(t-r/c) + \frac{\ell(\ell+1)}{2}\,\frac{c}{r}\,F^{(\ell-1)}(t-r/c)\bigg] + \mathcal{O}\left(\frac{1}{r^3}\right)\,.
\end{equation}
Injecting it into the field strength, it comes
\begin{subequations}
\begin{align}
&
F_{0i} 
= \frac{G\aem}{c^4\,r} \sum_{\ell\geq1} \frac{1}{c^\ell\,\ell !}\bigg[\perp_{ij}n_{L-1} \dE^{(\ell+1)}_{jL-1} - \frac{\ell}{c(\ell+1)}\,\varepsilon_{iab}n_{aL-1}\dB^{(\ell+1)}_{bL-1}\bigg]\nonumber\\
& \qquad\quad
+ \frac{G\aem}{c^3\,r^2} \sum_{\ell\geq1} \frac{1}{c^\ell\,\ell !}\bigg[\frac{\ell(\ell+1)}{2}\,n_{L-1}\dE^{(\ell)}_{iL-1} - \frac{(\ell+1)(\ell+2)}{2}\,n_{iL}\dE^{(\ell)}_{L} - \frac{\ell^2}{2c}\,\varepsilon_{iab}n_{aL-1}\dB^{(\ell)}_{bL-1}\bigg]\,,\\
&
F_{ij}
= \frac{G\aem}{c^4\,r} \sum_{\ell\geq1} \frac{1}{c^\ell\,\ell !}\bigg[n_{jL-1} \dE^{(\ell+1)}_{iL-1} -n_{iL-1} \dE^{(\ell+1)}_{jL-1} - \frac{\ell\,n_{aL-1}}{c(\ell+1)}\,\big(\varepsilon_{iab}n_j-\varepsilon_{jab}n_i\big)\dB^{(\ell+1)}_{bL-1}\bigg]\nonumber\\
& \qquad\quad
+ \frac{G\aem}{c^3\,r^2} \sum_{\ell\geq1} \frac{1}{c^\ell\,\ell !}\bigg[\frac{\ell(\ell+1)}{2}\bigg(n_{jL-1} \dE^{(\ell)}_{iL-1} -n_{iL-1} \dE^{(\ell)}_{jL-1}\bigg) \nonumber\\
& \hspace{5cm}
+\bigg(\frac{2\ell}{(\ell+1)c}\,\varepsilon_{ija}+\frac{\ell(\ell+2)}{2c}(\varepsilon_{iab}n_{jb}-\varepsilon_{jab}n_{ib})\bigg) n_{L-1}\dB^{(\ell)}_{aL-1}\nonumber\\
& \hspace{5cm}
+ \frac{\ell(\ell-1)}{(\ell+1)c}\bigg(\varepsilon_{iab}\dB^{(\ell)}_{jbL-2}-\varepsilon_{jab}\dB^{(\ell)}_{ibL-2}\bigg)n_{aL-2}\bigg]\,.
\end{align}
\end{subequations}
The flux~\eqref{eq_app_flux_GiEM} thus becomes
\begin{equation}
\mathcal{G}_\text{EM}^i 
=
\frac{G\aem}{4\pi c^3}\sum_{\ell,k}\frac{\ell+1}{c^{\ell+k}\,\ell!\,k!}\varepsilon_{iab} \Bigg[\dE_L^{(\ell)}\dE_{b K-1}^{(k+1)} + \frac{k\,\ell}{c^2(k+1)(\ell+1)}\dB_L^{(\ell)}\dB_{b K-1}^{(k+1)} \Bigg]\int\!\!\dd^2\Omega\, n_{a K-1}n_L\,.
\end{equation}
Using the integration formula
\begin{equation}
\varepsilon_{iab}\,\dE^{(\ell)}_L\dE^{(k+1)}_{bK-1}\,\int\!\!\dd^2 \Omega\,n_{aK-1}n_L =  \frac{4\pi\,\ell !}{(2\ell+1)!!}\,\delta_{k\ell}\,\varepsilon_{iab}\,\dE^{(\ell)}_{aL-1}\dE^{(\ell+1)}_{bL-1}\,,
\end{equation}
it comes
\begin{equation}
\mathcal{G}_\text{EM}^i 
=
\varepsilon_{iab}\,\sum_{\ell\geq 1}\frac{G\aem}{c^{3+2\ell}}\frac{\ell+1}{\ell!\,(2\ell+1)!!}\Bigg[\dE^{(\ell)}_{aL-1}\dE^{(\ell+1)}_{bL-1} + \frac{\ell^2}{c^2(\ell+1)^2}\,\dB^{(\ell)}_{aL-1}\dB^{(\ell+1)}_{bL-1}\bigg]\,.
\end{equation}
Under the replacement $\big(\dE_L^{(\ell)},\dB_L^{(\ell)}) \to (\dQ_L,\dH_L)$, this result gives Eqs.~\eqref{eq_flux_Gi_EM}, that we have used in this work.
Note however that when applying our formalism to a concrete case in Sec.~\ref{sec_toy_model}, we have only pushed the accuracy at NLO and the non-linearities did not contribute.
In this precise case, the ``source'' and ``full'' fluxes are naturally identical.

As discussed in Sec.~\ref{subsec_circ}, the balance equations~\eqref{eq_balance_eqs} are degenerate in the case of quasi-circular orbits.
Thus, the phase~\eqref{eq_phi_def} can also be obtained in terms of the balance equation for the angular momentum, \textit{i.e.}
\begin{equation}
\phi = - \frac{c^3}{Gm}\int\!\!\dd x\,\frac{x^{3/2}}{\mathcal{G}}\,\frac{\dd J}{\dd x}\,,
\end{equation}
where $\mathcal{G} = \vert \mathcal{G}^i\vert$ and $J = \vert J^i\vert$.
Implementing this formula, we recover exactly the previous result~\eqref{eq_phase_circ}.
This derivation can be seen as a check regarding the definition of the radiated EM angular momentum.
Indeed, the general definition of the GW angular momentum~\eqref{eq_flux_Gi_grav} in terms of the radiative moments was, of course, already known and its explicit value contains also magnetic contributions which combine to those of the EM flux to give the expected result.
The fact that both derivations agree thus strongly suggests that the adopted definition is correct.
As a sidenote, it is also remarkable that the link between the coefficients of $\mathcal{F}_\text{EM}$ and $\mathcal{G}_\text{EM}$ is the same as the proportionality coefficients of those between $\mathcal{F}_\text{GW}$ and $\mathcal{G}_\text{GW}$.

\section{Lengthy expressions}\label{app_lengthy_expr}

All the lengthy results presented hereafter are collected in the ancillary file~\cite{AncFile}.

\subsection{Expressions of the fluxes for generic orbits}\label{app_expr_flux}

This appendix presents the explicit expressions of the fluxes computed in Sec~\ref{subsec_Flux_toymodel}. We recall that the total energy flux for arbitrary planar motion has been decomposed as
\begin{equation}\label{eq_energy_flux_gen}
\mathcal{F} = \mathcal{F}_\text{GW,pp} + \mathcal{F}_\text{GW,mag} + \mathcal{F}_\text{EM}\,.
\end{equation}
The (instantaneous) point particle sector, $\mathcal{F}_\text{GW,pp}$, can be found up to 3PN in Eq.~(5.2) of~\cite{Arun:2007sg}, and is given at 1PN by
\begin{equation}
\begin{aligned}
\mathcal{F}_\text{GW,pp} 
&
= \frac{32G^3m^4\nu^2}{5c^5r^4} \Bigg\lbrace
r^2\dot{\phi}^2 + \frac{\dot{r}^2}{12}\\
& \hspace{2.8cm}
+ \frac{1}{c^2}\Bigg[
\bigg(\frac{785}{336}- \frac{71\nu}{28}\bigg)r^4\dot{\phi}^4
- \bigg(\frac{117}{28}- \frac{45\nu}{14}\bigg)r^2\dot{r}^2\dot{\phi}^2
- \bigg(\frac{8}{21}- \frac{3\nu}{14}\bigg)\dot{r}^4\\
& \hspace{4cm}
- \bigg(\frac{170}{21}- \frac{10\nu}{21}\bigg)Gmr\dot{\phi}^2
+ \bigg(\frac{9}{14}+ \frac{5\nu}{42}\bigg)\frac{Gm\dot{r}^2}{r}
+ \frac{1-4\nu}{21}\frac{G^2m^2}{r^2}
\Bigg]\Bigg\rbrace\,.
\end{aligned}
\end{equation}
As for the EM contributions, they read at NLO
\begin{subequations}
\begin{align}
\mathcal{F}_\text{GW,mag}
& 
\nonumber
= - \frac{192G^5m^6\nu\aem \tilde{\mu}_1\tilde{\mu}_2}{5c^9r^6}\Bigg\lbrace
r^2\dot{\phi}^2 - \frac{\dot{r}^2}{12}\\
& \hspace{5cm}\nonumber
+ \frac{1}{c^2}\Bigg[
\bigg(\frac{277}{112}- \frac{569\nu}{84}\bigg)r^4\dot{\phi}^4
- \bigg(\frac{1159}{168}- \frac{1373\nu}{84}\bigg)r^2\dot{r}^2\dot{\phi}^2\\
& \hspace{6cm}\nonumber
+ \bigg(\frac{17}{42}- \frac{37\nu}{42}\bigg)\dot{r}^4
- \bigg(\frac{1331}{168}-\frac{27\nu}{7}\bigg)Gmr\dot{\phi}^2\\
& \hspace{6cm}
+ \bigg(\frac{17}{18} - \frac{109\nu}{84}\bigg)\frac{Gm\dot{r}^2}{r}
+ \frac{1-4\nu}{14}\frac{G^2m^2}{r^2}
\Bigg]\Bigg\rbrace\,,\\
\mathcal{F}_\text{EM}
& 
\nonumber
= \frac{4G^5m^6\nu^2\aem \tilde{\mu}_-^2}{15c^9r^6}\Bigg\lbrace
r^2\dot{\phi}^2 +4\dot{r}^2\\
& \hspace{4cm}\nonumber
+ \frac{1}{c^2}\Bigg[
\bigg(\frac{397}{28}- \frac{465\nu}{14}\bigg)r^4\dot{\phi}^4
+ \bigg(\frac{1177}{14}- \frac{906\nu}{7}\bigg)r^2\dot{r}^2\dot{\phi}^2\\
& \hspace{5cm}\nonumber
- \bigg(\frac{128}{7}- \frac{324\nu}{7}\bigg)\dot{r}^4
- \bigg(\frac{71}{14} + \frac{45\nu}{7}\bigg)Gmr\dot{\phi}^2\\
& \hspace{5cm}\nonumber
- \bigg(\frac{488}{7} - \frac{648\nu}{7}\bigg)\frac{Gm\dot{r}^2}{r}
+ 8\frac{1-4\nu}{7}\frac{G^2m^2}{r^2}
\Bigg]\Bigg\rbrace\\
& \nonumber\quad
+ \frac{64G^5m^6\nu^2 \delta \aem \tilde{\mu}_+\tilde{\mu}_-}{105c^{11}r^6}\Bigg[
\frac{33}{4}r^4\dot{\phi}^4+\frac{471}{16}r^2\dot{r}^2\dot{\phi}^2-6\dot{r}^4
+ \frac{9}{8}Gmr\dot{\phi}^2 - \frac{123}{8}\frac{Gm\dot{r}^2}{r} + \frac{G^2m^2}{r^2}\Bigg]\\
& \quad
+ \frac{32G^5m^6\nu^2\aem \tilde{\mu}_+^2}{105c^{11}r^6}\Bigg[
10r^4\dot{\phi}^4+\frac{83}{4}r^2\dot{r}^2\dot{\phi}^2+4\dot{r}^4
+ Gmr\dot{\phi}^2 + \frac{4Gm\dot{r}^2}{r} + \frac{G^2m^2}{r^2}\Bigg]
\,.
\end{align}
\end{subequations}
Note that we have written $\mathcal{F}_\text{GW,mag}$ in terms of $\tilde{\mu}_{1,2}$ instead of $\tilde{\mu}_\pm$, for the sake of compactness.
Indeed, it comes
\begin{equation}
\tilde{\mu}_1\tilde{\mu}_2 = \nu\left(\tilde{\mu}_+^2+\delta\,\tilde{\mu}_+\tilde{\mu}_--\nu\tilde{\mu}_-^2\right)\,.
\end{equation}
Similarly to the energy flux, the total angular momentum flux is split according to
\begin{equation}\label{eq_ang_flux_gen}
\mathcal{G}^i = \bigg[\mathcal{G}_\text{GW,pp} + \mathcal{G}_\text{GW,mag} + \mathcal{G}_\text{EM}\bigg]l^i\,,
\end{equation}
where the (instantaneous) point-particle sector, $\mathcal{G}_\text{GW,pp}$, can be found at 3PN \emph{e.g.} in Eq.~(3.2) of~\cite{Arun:2009mc}. 
At the required order, the contributions read
\begin{subequations}
\begin{align}
\mathcal{G}_\text{GW,pp} 
&
\nonumber
= \frac{16G^2m^3\nu^2\dot{\phi}}{5c^5r} \Bigg\lbrace
r^2\dot{\phi}^2 - \frac{\dot{r}^2}{2} + \frac{Gm}{r}\\
& \hspace{2.8cm}\nonumber
+ \frac{1}{c^2}\Bigg[
\bigg(\frac{307}{168}- \frac{137\nu}{42}\bigg)r^4\dot{\phi}^4
+ \bigg(\frac{85}{84}+ \frac{283\nu}{84}\bigg)r^2\dot{r}^2\dot{\phi}^2
+ \bigg(\frac{37}{42}+ \frac{17\nu}{84}\bigg)\dot{r}^4\\
& \hspace{4cm}
- \bigg(\frac{29}{21}+ \frac{95\nu}{42}\bigg)Gmr\dot{\phi}^2
+ \bigg(\frac{64}{21}+ \frac{\nu}{12}\bigg)\frac{Gm\dot{r}^2}{r}
- \bigg(\frac{745}{84}- \frac{\nu}{42}\bigg)\frac{G^2m^2}{r^2}
\Bigg]\Bigg\rbrace\,,\\
\mathcal{G}_\text{GW,mag}
& 
\nonumber
= - \frac{48G^4m^5\nu\aem \tilde{\mu}_1\tilde{\mu}_2\,\dot{\phi}}{5c^9r^3}\Bigg\lbrace
r^2\dot{\phi}^2 - \frac{3\dot{r}^2}{2} + \frac{2Gm}{r}\\
& \hspace{5cm}\nonumber
+ \frac{1}{c^2}\Bigg[
\bigg(\frac{323}{168}- \frac{22\nu}{3}\bigg)r^4\dot{\phi}^4
- \bigg(\frac{167}{14}- \frac{1221\nu}{28}\bigg)r^2\dot{r}^2\dot{\phi}^2\\
& \hspace{6cm}\nonumber
+ \bigg(\frac{27}{7}- \frac{1181\nu}{84}\bigg)\dot{r}^4
+ \bigg(\frac{125}{84}- \frac{229\nu}{21}\bigg)Gmr\dot{\phi}^2\\
& \hspace{6cm}
+ \bigg(\frac{215}{84} + \frac{227\nu}{42}\bigg)\frac{Gm\dot{r}^2}{r}
- \bigg(\frac{356}{21} - \frac{116\nu}{21}\bigg)\frac{G^2m^2}{r^2}
\Bigg]\Bigg\rbrace\,,\\
\mathcal{G}_\text{EM}
& 
\nonumber
= \frac{4G^5m^6\nu^2\aem \tilde{\mu}_-^2\,\dot{\phi}}{15c^9r^4}\Bigg\lbrace
1+ \frac{1}{c^2}\Bigg[
\bigg(\frac{223}{14}- \frac{282\nu}{7}\bigg)r^2\dot{\phi}^2
- \bigg(\frac{151}{28}+ \frac{207\nu}{14}\bigg)\dot{r}^2\\
&\hspace{6cm}\nonumber-\bigg(\frac{159}{28}+ \frac{55\nu}{14}\bigg)\frac{G m}{r}
\Bigg]\Bigg\rbrace\\
& \nonumber\quad
+ \frac{614G^5m^6\nu^2 \delta \aem \tilde{\mu}_+\tilde{\mu}_-\,\dot{\phi}}{105c^{11}r^4}\Bigg[r^2\dot{\phi}^2- \frac{77\dot{r}^2}{307} + \frac{25Gm}{307r}\Bigg]\\
& \quad
+ \frac{16G^5m^6\nu^2 \aem \tilde{\mu}_+^2\,\dot{\phi}}{5c^{11}r^4}\Bigg[r^2\dot{\phi}^2- \frac{\dot{r}^2}{14} + \frac{Gm}{7r}\Bigg]
\,.
\end{align}
\end{subequations}

\subsection{Expressions of the gravitational modes for generic orbits}\label{app_modes}

This appendix presents the instantaneous gravitational modes computed in Sec~\ref{subsec_Flux_toymodel}.
For the sake of simplicity, we single out the dependence on the gravitational phase $\phi$ and rescale the modes $h_{\ell \dm}$~\eqref{eq_mode_hlm2ULVL} as
\begin{equation}\label{eq_mode_Hlm_gen}
h_{\ell \dm}
= \frac{8 G m \nu}{c^4 R} \,\sqrt{\frac{\pi}{5}} \,\mathcal{H}_{\ell \dm} \, \de^{-\di \dm \phi}\,,
\end{equation}
and further split $\mathcal{H}_{\ell \dm} = \mathcal{H}_{\ell \dm}^\text{pp} + \mathcal{H}_{\ell \dm}^\text{EM}$.
Moreover, those coefficients obey $\mathcal{H}_{\ell,-\dm} = (-)^\ell \,\overline{\mathcal{H}}_{\ell \dm}$, so we only present the ones with a positive $\dm$.
The point-particle sector, $\mathcal{H}_{\ell \dm}^\text{pp}$ can be found \emph{e.g.} in~\cite{Henry:2023tka}, and we don't replicate it here.
As for the EM corrections, the instantaneous modes read to the relevant order
\begin{subequations}
\begin{align}
&
\mathcal{H}_{22}^\text{EM} =
\nonumber
- \frac{3}{2}\frac{G^3m^3\aem\tilde{\mu}_1\tilde{\mu}_2}{c^4r^3\,\nu}\Bigg\lbrace 1+\frac{1}{c^2}\Bigg[
\bigg(\frac{1}{21}-\frac{8\nu}{7}\bigg)r^2\dot{\phi}^2
-\di \bigg(\frac{73}{21}-\frac{73\nu}{7}\bigg)r\,\dot{r}\dot{\phi}\\
& \hspace{7cm}
- \bigg(\frac{9}{14}-\frac{24\nu}{7}\bigg)\dot{r}^2
- \bigg(\frac{187}{42}-\frac{20\nu}{7}\bigg)\frac{Gm}{r}\Bigg]\Bigg\rbrace\,,\\
&
\mathcal{H}_{21}^\text{EM} =
- \di\,\frac{G^3m^3\aem\tilde{\mu}_1\tilde{\mu}_2\,\delta}{c^5r^2\,\nu}\,\dot{\phi}\,,\\
&
\mathcal{H}_{20}^\text{EM} =
\sqrt{\frac{3}{2}}\frac{G^3m^3\aem\tilde{\mu}_1\tilde{\mu}_2}{c^4r^3\,\nu}\Bigg\lbrace 1+\frac{1}{c^2}\Bigg[
\bigg(\frac{3}{7}-\frac{16\nu}{7}\bigg)r^2\dot{\phi}^2
- \bigg(\frac{9}{14}-\frac{24\nu}{7}\bigg)\dot{r}^2
- \bigg(\frac{187}{42}-\frac{20\nu}{7}\bigg)\frac{Gm}{r}\Bigg]\Bigg\rbrace\,,\\
&
\mathcal{H}_{33}^\text{EM} =
\frac{\di}{4}\sqrt{\frac{105}{2}}\frac{G^3m^3\aem\tilde{\mu}_1\tilde{\mu}_2\,\delta}{c^5r^3\,\nu}\bigg(r\dot{\phi}+ \frac{2}{7}\,\di\,\dot{r}\bigg)\,,\\
&
\mathcal{H}_{32}^\text{EM} =
- \sqrt{\frac{5}{7}}\frac{G^3m^3\aem\tilde{\mu}_1\tilde{\mu}_2}{c^6r^2\,\nu}\big(1-3\nu\big)\,\bigg(r\dot{\phi}- \frac{\di \,\dot{r}}{4}\bigg)\,\dot{\phi}\,,\\
&
\mathcal{H}_{31}^\text{EM} =
-\frac{\di}{4}\sqrt{\frac{7}{2}}\frac{G^3m^3\aem\tilde{\mu}_1\tilde{\mu}_2\,\delta}{c^5r^3\,\nu}\bigg(r\dot{\phi}+ \frac{6}{7}\,\di\,\dot{r}\bigg)\,,\\
&
\mathcal{H}_{30}^\text{EM} =
-\frac{\di}{2} \sqrt{\frac{3}{14}}\frac{G^3m^3\aem\tilde{\mu}_1\tilde{\mu}_2}{c^6r^2\,\nu}\big(1-3\nu\big)\,\dot{r}\,\dot{\phi}\,,\\
&
\mathcal{H}_{44}^\text{EM} =
\frac{53}{24}\sqrt{\frac{5}{7}}\frac{G^3m^3\aem\tilde{\mu}_1\tilde{\mu}_2}{c^6r^3\,\nu}\big(1-3\nu\big)\,\bigg[r^2\dot{\phi}^2 + \frac{26\,\di}{53}r\,\dot{r}\dot{\phi}- \frac{8}{53}\dot{r}^2+ \frac{12}{53}\frac{Gm}{r}\bigg]\,,\\
&
\mathcal{H}_{42}^\text{EM} =
-\frac{5\sqrt{5}}{84}\frac{G^3m^3\aem\tilde{\mu}_1\tilde{\mu}_2}{c^6r^3\,\nu}\big(1-3\nu\big)\,\bigg[r^2\dot{\phi}^2 + \frac{13\,\di}{5}r\,\dot{r}\dot{\phi}- \frac{8}{5}\dot{r}^2+ \frac{12}{5}\frac{Gm}{r}\bigg]\,,\\
&
\mathcal{H}_{40}^\text{EM} =
-\frac{11}{28\sqrt{2}}\frac{G^3m^3\aem\tilde{\mu}_1\tilde{\mu}_2}{c^6r^3\,\nu}\big(1-3\nu\big)\,\bigg[r^2\dot{\phi}^2 + \frac{8}{11}\dot{r}^2 - \frac{12}{11}\frac{Gm}{r}\bigg]\,.
\end{align}
\end{subequations}

\subsection{Quasi-Keplerian parameters}\label{app_qK_par}

The parameters entering the quasi-Keplerian representation of the motion~\eqref{eq_qK_parametrisation} are given herebelow in terms of the two dimensionless invariants linked to the (unreduced) energy $E$ and norm of the angular momentum $J = \vert J^i \vert$
\begin{equation}
\varepsilon \equiv \frac{-2E}{m\,\nu\,c^2}
\qquad\text{and}\qquad
j \equiv \frac{-2E\,J^2}{G^2m^5\,\nu^3}\,,
\end{equation}
where we recall that we deal with bounded trajectories, thus $E <0$.
While the energy parameter is a 1PN quantity, the second parameter is Newtonian and is such that the eccentricities are simply $e_\star = \sqrt{1-j} + \mathcal{O}(c^{-2})$.
It relates to the usual definition $h = J/(Gm^2\nu)$ by $j = -2h^2E/(m\nu)$.
At the required order, we obtain
\begin{subequations}
\begin{align}
&\label{eq_qK_ar}
a_r =
\frac{Gm}{c^2\,\varepsilon}\Bigg\lbrace
1- \frac{7-\nu}{4}\varepsilon
+ \frac{\aem\,\tilde{\mu}_1\tilde{\mu}_2}{j\,\nu}\,\varepsilon^2\,\bigg[1 + \frac{12+\nu}{j}\,\varepsilon\bigg]\Bigg\rbrace\,,\\
&\label{eq_qK_n}
n = \frac{c^3}{G m}\,\varepsilon^{3/2}\,\Bigg\lbrace1- \frac{15-\nu}{8}\varepsilon
+ \big(7-2\nu\big)\frac{\aem\,\tilde{\mu}_1\tilde{\mu}_2}{j^{3/2}\nu}
\,\varepsilon^3\Bigg\rbrace\,,\\
&\label{eq_qK_K}
K = 1 + \frac{3}{j}\,\varepsilon
- \frac{3\aem\,\tilde{\mu}_1\tilde{\mu}_2}{j^2\,\nu}\,\varepsilon^2\bigg[1 + \bigg(\frac{45}{2j}- \frac{7}{2}+\nu\bigg)\,\varepsilon\bigg]\,,\\
&\label{eq_qK_er}
e_r^2 =
1 - j + \bigg(6-\nu - \frac{15-5\nu}{4}\,j \bigg)\varepsilon
\nonumber\\
&
\hspace{2cm}
- \frac{2\aem\,\tilde{\mu}_1\tilde{\mu}_2}{j\,\nu}\,\varepsilon^2\,\bigg[2-j + \bigg( \frac{24+2\nu}{j}- \frac{11+7\nu}{2} - \frac{11-3\nu}{2}\,j \bigg)\varepsilon\bigg]
\,,\\
&\label{eq_qK_et}
e_t^2 =
1 - j - \bigg(2-2\nu - \frac{17-7\nu}{4}\,j \bigg)\varepsilon
\nonumber\\
&
\hspace{2cm}
- \frac{2\aem\,\tilde{\mu}_1\tilde{\mu}_2}{j\,\nu}\,\varepsilon^2\,\bigg[1 + \bigg( \frac{12+\nu}{j} - \frac{7-2\nu}{j^{1/2}}- \frac{37-7\nu}{4} +(7-2\nu)j^{1/2} \bigg)\varepsilon\bigg]
\,,\\
&\label{eq_qK_ephi}
e_\phi^2 =
1 - j + \bigg(6-\frac{15-\nu}{4}\,j \bigg)\varepsilon
\nonumber\\
&
\hspace{2cm}
- \frac{2\aem\,\tilde{\mu}_1\tilde{\mu}_2}{j\,\nu}\,\varepsilon^2\,\bigg[3-2j + \bigg( \frac{336-13\nu}{8\,j} -  \frac{63-16\nu}{4} -\frac{88+9\nu}{8}\,j\bigg)\varepsilon\bigg]
\,,\\
&\label{eq_qK_Fvu}
f_{v-u} = -\big(7-2\nu)\frac{\aem\,\tilde{\mu}_1\tilde{\mu}_2}{j^{3/2}\,\nu}\,\varepsilon^3\,,\\
&\label{eq_qK_Fv}
f_{v} =\frac{3\aem\,\tilde{\mu}_1\tilde{\mu}_2}{2}\,\frac{\sqrt{1-j}}{j^{3/2}}\,\varepsilon^3\,,\\
&\label{eq_qK_G2v}
g_{2v} =-\frac{3-4\nu}{4}\frac{\aem\,\tilde{\mu}_1\tilde{\mu}_2}{\nu}\frac{1-j}{j^3}\,\varepsilon^3\,,\\
&\label{eq_qK_G3v}
g_{3v} =\frac{\aem\,\tilde{\mu}_1\tilde{\mu}_2}{8}\frac{(1-j)^{3/2}}{j^3}\,\varepsilon^3\,.
\end{align}
\end{subequations}
The 1PN point-particle sector of those expressions naturally agrees with~\cite{DamourpK}.
For the sake of completeness, we also display the expression of $x$~\eqref{eq_xdef_qK} at the sought order
\begin{equation}\label{eq_qK_x}
x = \varepsilon \,\Bigg\lbrace1 + \bigg[\frac{2}{j}- \frac{15-\nu}{12} \bigg]\,\varepsilon - \frac{2\aem\,\tilde{\mu}_1\tilde{\mu}_2}{j^2\,\nu}\,\varepsilon^2\bigg[1+ \bigg(\frac{43}{2j}- \frac{57-13\nu}{12}- \frac{7-2\nu}{3}j^{1/2} \bigg)\varepsilon\bigg]\Bigg\rbrace\,.
\end{equation}

\subsection{Expressions of the fluxes for eccentric orbits}\label{app_qK_flux}

In the spirit of Eqs.~(8.8) of~\cite{Arun:2007sg} and~(4.10) of~\cite{Arun:2009mc}, respectively presenting the averaged instantaneous energy and angular momentum fluxes in a system of modified harmonic coordinates, we split at the NLO
\begin{subequations}\label{eq_qK_FG_avg}
\begin{align}
&
\left\langle \mathcal{F}\right\rangle =
\frac{32\,\nu^2c^5}{5G}x^5\Bigg[\mathcal{F}_\text{N} + x\,\mathcal{F}_{1\text{PN}} + \aem\,x^2\bigg(\mathcal{F}_\text{mag,LO}+x\,\mathcal{F}_\text{mag,NLO}\bigg)\Bigg]\,,\label{eq_qK_F_avg}\\
&
\left\langle \mathcal{G}\right\rangle =
\frac{4}{5}\,m\nu^2c^2\,x^{7/2}\Bigg[\mathcal{G}_\text{N} + x\,\mathcal{G}_{1\text{PN}} + \aem\,x^2\bigg(\mathcal{G}_\text{mag,LO}+x\,\mathcal{G}_\text{mag,NLO}\bigg)\Bigg]\,,
\,.
\end{align}
\end{subequations}
with (let us emphasize that the following expressions are exact, in the sense that they have not been truncated in powers of $e_t$)
\begin{subequations}
\begin{align}
\mathcal{F}_\text{N}
&
=
\frac{1}{(1-e_t^2)^{7/2}}\Bigg[1 + \frac{73\,e_t^2}{24} + \frac{37\,e_t^4}{96}\Bigg]\,,\\
\mathcal{F}_{1\text{PN}} 
&
=
\frac{1}{(1-e_t^2)^{9/2}}\Bigg[- \frac{1247}{336} - \frac{35\nu}{12} + \bigg(\frac{10475}{672} - \frac{1081\nu}{36}\bigg)e_t^2
\nonumber\\
& \hspace{3cm} + \bigg(\frac{10043}{384} - \frac{311\nu}{12}\bigg)e_t^4 + \bigg(\frac{2179}{1792} - \frac{851\nu}{576}\bigg)e_t^6\Bigg]\,,\\
\mathcal{F}_\text{mag,LO}
&
=
\frac{\tilde{\mu}_+^2+\delta\,\tilde{\mu}_+\tilde{\mu}_--\nu\tilde{\mu}_-^2}{(1-e_t^2)^{11/2}}\Bigg[-4- \frac{101\,e_t^2}{2}-\frac{1999\,e_t^4}{48}- \frac{59\,e_t^6}{32}\Bigg]
\nonumber\\
&\quad
+ \frac{\tilde{\mu}_-^2}{(1-e_t^2)^{11/2}}\bigg[\frac{1}{24}+ \frac{19\,e_t^2}{48}+\frac{23\,e_t^4}{64}+ \frac{3\,e_t^6}{128}\Bigg]
\,,\\
\mathcal{F}_\text{mag,NLO}
&
=\frac{\tilde{\mu}_+^2+\delta\,\tilde{\mu}_+\tilde{\mu}_--\nu\tilde{\mu}_-^2}{(1-e_t^2)^{13/2}}\Bigg[
\frac{663}{56} + \frac{47\nu}{2} - \bigg(\frac{28579}{84} - \frac{4903\nu}{9}\bigg)e_t^2- \bigg(\frac{138987}{112} - \frac{50419\nu}{36}\bigg)e_t^4 
\nonumber\\
& \hspace{5cm} - \bigg(\frac{318767}{672} - \frac{157811\nu}{288}\bigg)e_t^6 - \bigg(\frac{69525}{7168} - \frac{1537\nu}{96}\bigg)e_t^8\Bigg]
\nonumber\\
& \quad
+
\frac{\tilde{\mu}_+^2+\delta\,\tilde{\mu}_+\tilde{\mu}_--\nu\tilde{\mu}_-^2}{(1-e_t^2)^6}\Bigg[- \frac{49}{3}+ \frac{14\nu}{3} - \bigg(\frac{8995}{72} - \frac{1285\nu}{36}\bigg)e_t^2
\nonumber\\
& \hspace{5cm} - \bigg(\frac{7091}{96} - \frac{1013\nu}{48}\bigg)e_t^4 - \bigg(\frac{259}{144} - \frac{37\nu}{72}\bigg)e_t^6\Bigg]
\nonumber\\
& \quad
+ \frac{\delta\,\tilde{\mu}_+\tilde{\mu}_-}{(1-e_t^2)^{13/2}}\Bigg[\frac{5}{12} + \frac{259\,e_t^2}{48}+\frac{1963\,e_t^4}{192}+ \frac{1441\,e_t^6}{384}+ \frac{141\,e_t^8}{1024}\Bigg]
\nonumber\\
& \quad
+ \frac{\tilde{\mu}_-^2}{(1-e_t^2)^{13/2}}\Bigg[
\frac{19}{32} - \frac{191\nu}{144} + \bigg(\frac{93}{8} - \frac{2627\nu}{144}\bigg)e_t^2 + \bigg(\frac{10915}{384} - \frac{10333\nu}{288}\bigg)e_t^4 
\nonumber\\
& \hspace{4cm} + \bigg(\frac{1079}{96} - \frac{5099\nu}{384}\bigg)e_t^6 + \bigg(\frac{1533}{4096} - \frac{983\nu}{2048}\bigg)e_t^8\Bigg]
\,,
\end{align}
\end{subequations}
and
\begin{subequations}
\begin{align}
\mathcal{G}_\text{N}
&
=
\frac{8+7e_t^2}{(1-e_t^2)^2}\,,\\
\mathcal{G}_{1\text{PN}} 
&
=
\frac{1}{(1-e_t^2)^3}\Bigg[- \frac{1247}{42} - \frac{70\nu}{3} + \bigg(\frac{3019}{42} - \frac{335\nu}{3}\bigg)e_t^2 + \bigg(\frac{8399}{336} - \frac{275\nu}{12}\bigg)e_t^4 \Bigg]\,,\\
\mathcal{G}_\text{mag,LO}
&
=
\frac{\tilde{\mu}_+^2+\delta\,\tilde{\mu}_+\tilde{\mu}_--\nu\tilde{\mu}_-^2}{(1-e_t^2)^4}\Bigg[-32- 174\,e_t^2-\frac{63\,e_t^4}{2}\Bigg]
+ \frac{\tilde{\mu}_-^2}{(1-e_t^2)^4}\bigg[\frac{1}{3}+ e_t^2+\frac{e_t^4}{8}\Bigg]
\,,\\
\mathcal{G}_\text{mag,NLO}
&
=\frac{\tilde{\mu}_+^2+\delta\,\tilde{\mu}_+\tilde{\mu}_--\nu\tilde{\mu}_-^2}{(1-e_t^2)^5}\Bigg[
\frac{271}{7} + 204\nu - \bigg(\frac{63443}{42} - \frac{7192\nu}{3}\bigg)e_t^2 
\nonumber\\
& \hspace{5cm} - \bigg(\frac{363301}{168} - \frac{15529\nu}{6}\bigg)e_t^4 - \bigg(\frac{17091}{112} -  206\nu\bigg)e_t^6\Bigg]
\nonumber\\
& \quad
-
\frac{\tilde{\mu}_+^2+\delta\,\tilde{\mu}_+\tilde{\mu}_--\nu\tilde{\mu}_-^2}{(1-e_t^2)^{11/2}}\,\frac{2(7-2\nu)(16+41\,e_t^2-60\,e_t^4-7\,e_t^6 )}{3}
\nonumber\\
& \quad
+ \frac{\delta\,\tilde{\mu}_+\tilde{\mu}_-}{(1-e_t^2)^5}\Bigg[\frac{10}{3} + \frac{145\,e_t^2}{6}+\frac{35\,e_t^4}{2}+ \frac{15\,e_t^6}{16}\Bigg]
\nonumber\\
& \quad
+ \frac{\tilde{\mu}_-^2}{(1-e_t^2)^5}\Bigg[
\frac{19}{4} - \frac{191\nu}{18} + \bigg(\frac{385}{8} - \frac{2867\nu}{36}\bigg)e_t^2
\nonumber\\
& \hspace{3cm}  +  \bigg(\frac{1263}{32} - \frac{2845\nu}{48}\bigg)e_t^4  + \bigg(\frac{131}{64} - \frac{311\nu}{96}\bigg)e_t^6 \Bigg]
\,.
\end{align}
\end{subequations}
Naturally, the point-particle sectors at 1PN agree with Eqs~(8.9) of~\cite{Arun:2007sg} and~(4.11) of~\cite{Arun:2009mc}. 

\subsection{Time evolution of the orbital parameters}\label{app_qK_orbital}

Using the chain rule~\eqref{eq_dedt_chain_rule} together with Eq.~\eqref{eq_qK_et} and the averaged fluxes~\eqref{eq_qK_FG_avg}, we find the evolution of the time-eccentricity as
\begin{equation}\label{eq_detdt_split}
\left\langle \frac{\dd e_t}{\dd t}\right\rangle = - \frac{c^3\,e_t\,\nu}{Gm}\,x^4\,\Bigg[\mathcal{E}_\text{N} + x\,\mathcal{E}_{1\text{PN}} + \aem\,x^2\bigg(\mathcal{E}_\text{mag,LO}+x\,\mathcal{E}_\text{mag,NLO}\bigg)\Bigg]\,,
\end{equation}
where
\begin{subequations}
\begin{align}
\mathcal{E}_\text{N}
&
=
\frac{1}{(1-e_t^2)^{5/2}}\Bigg[\frac{304}{15} + \frac{121\,e_t^2}{15}\Bigg]\,,\\
\mathcal{E}_{1\text{PN}} 
&
=
\frac{1}{(1-e_t^2)^{7/2}}\Bigg[- \frac{939}{35} - \frac{4084\nu}{45} + \bigg(\frac{29917}{105} - \frac{7753\nu}{30}\bigg)e_t^2 + \bigg(\frac{13929}{280} - \frac{1664\nu}{45}\bigg)e_t^4 \Bigg]\,,\\
\mathcal{E}_\text{mag,LO}
&
=
\frac{\tilde{\mu}_+^2+\delta\,\tilde{\mu}_+\tilde{\mu}_--\nu\tilde{\mu}_-^2}{(1-e_t^2)^{9/2}}\Bigg[-\frac{3752}{15}- \frac{1190\,e_t^2}{3}-37\,e_t^4\Bigg]
+ \frac{\tilde{\mu}_-^2}{(1-e_t^2)^{9/2}}\bigg[2+ 3\,e_t^2+\frac{e_t^4}{4}\Bigg]
\,,\\
\mathcal{E}_\text{mag,NLO}
&
=\frac{\tilde{\mu}_+^2+\delta\,\tilde{\mu}_+\tilde{\mu}_--\nu\tilde{\mu}_-^2}{(1-e_t^2)^{11/2}}\Bigg[
-\frac{16220}{21} + \frac{72916\nu}{45} - \bigg(\frac{1778741}{210} - \frac{27724\nu}{3}\bigg)e_t^2
\nonumber\\
& \hspace{5cm} - \bigg(\frac{267145}{42} - \frac{543343\nu}{90}\bigg)e_t^4  - \bigg(\frac{321277}{1120} - \frac{3099\nu}{10}\bigg)e_t^6 \Bigg]
\nonumber\\
& \quad
-
\frac{\tilde{\mu}_+^2+\delta\,\tilde{\mu}_+\tilde{\mu}_--\nu\tilde{\mu}_-^2}{(1-e_t^2)^5}\,\frac{(7-2\nu)(3248+6891\,e_t^2+61\,e_t^4)}{45}
\nonumber\\
& \quad
+ \frac{\delta\,\tilde{\mu}_+\tilde{\mu}_-}{(1-e_t^2)^{11/2}}\Bigg[\frac{268}{15} + \frac{2123\,e_t^2}{30}+\frac{559\,e_t^4}{15}+ \frac{261\,e_t^6}{160}\Bigg]
\nonumber\\
& \quad
+ \frac{\tilde{\mu}_-^2}{(1-e_t^2)^{11/2}}\Bigg[
\frac{1013}{30} - \frac{872\nu}{15} + \bigg(\frac{7659}{40} - \frac{14677\nu}{60}\bigg)e_t^2 
\nonumber\\
& \hspace{4cm} + \bigg(\frac{27527}{240} - \frac{2683\nu}{20}\bigg)e_t^4  + \bigg(\frac{641}{128} - \frac{1927\nu}{320}\bigg)e_t^6 \Bigg]
\,.
\end{align}
\end{subequations}
Let us emphasize again that the expressions presented in this section are exact, in the sense that they have not been truncated in powers of $e_t$.
Similarly, the semi-major axis decays as
\begin{equation}\label{eq_dardt_split}
\left\langle \frac{\dd a_r}{\dd t}\right\rangle = \nu \,c\,x^3\,\Bigg[\mathcal{A}_\text{N} + x\,\mathcal{A}_{1\text{PN}} + \aem\,x^2\bigg(\mathcal{A}_\text{mag,LO}+x\,\mathcal{A}_\text{mag,NLO}\bigg)\Bigg]\,,
\end{equation}
where
\begin{subequations}
\begin{align}
\mathcal{A}_\text{N}
&
=
\frac{1}{(1-e_t^2)^{7/2}}\Bigg[- \frac{64}{5} - \frac{584\,e_t^2}{15} - \frac{74\,e_t^4}{15}\Bigg]\,,\\
\mathcal{A}_{1\text{PN}} 
&
=
\frac{1}{(1-e_t^2)^{9/2}}\Bigg[\frac{2972}{105} + \frac{176\nu}{5} - \bigg(\frac{30442}{105} - 380\nu\bigg)e_t^2
\nonumber\\
& \hspace{3cm} - \bigg(\frac{879}{2} - \frac{1687\nu}{5}\bigg)e_t^4 - \bigg(\frac{11717}{420} - \frac{296\nu}{15}\bigg)e_t^6\Bigg]\,,\\
\mathcal{A}_\text{mag,LO}
&
=
\frac{\tilde{\mu}_+^2+\delta\,\tilde{\mu}_+\tilde{\mu}_--\nu\tilde{\mu}_-^2}{(1-e_t^2)^{11/2}}\Bigg[\frac{576}{5} + \frac{11816\,e_t^2}{15} + \frac{2716\,e_t^4}{5} + \frac{174\,e_t^6}{5}\Bigg]
\nonumber\\
&\quad
- \frac{\tilde{\mu}_-^2}{(1-e_t^2)^{11/2}}\bigg[\frac{8}{15} + \frac{76\,e_t^2}{15}+\frac{23\,e_t^4}{5}+ \frac{3\,e_t^6}{10}\Bigg]
\,,\\
\mathcal{A}_\text{mag,NLO}
&
=\frac{\tilde{\mu}_+^2+\delta\,\tilde{\mu}_+\tilde{\mu}_--\nu\tilde{\mu}_-^2}{(1-e_t^2)^{13/2}}\Bigg[
\frac{1156}{3} - \frac{2192\nu}{5} + \bigg(\frac{171160}{21} - \frac{78572\nu}{9}\bigg)e_t^2 + \bigg(\frac{125977}{6} - \frac{884126\nu}{45}\bigg)e_t^4 
\nonumber\\
& \hspace{5cm} + \bigg(\frac{55471}{7} - \frac{109891\nu}{15}\bigg)e_t^6 + \bigg(\frac{459559}{1680} - \frac{4103\nu}{15}\bigg)e_t^8\Bigg]
\nonumber\\
& \quad
+ \frac{\tilde{\mu}_+^2+\delta\,\tilde{\mu}_+\tilde{\mu}_--\nu\tilde{\mu}_-^2}{(1-e_t^2)^6}\,\frac{2(7-2\nu)(96+1452\,e_t^2+1353\,e_t^4+74\,e_t^6)}{15}
\nonumber\\
& \quad
- \frac{\delta\,\tilde{\mu}_+\tilde{\mu}_-}{(1-e_t^2)^{13/2}}\Bigg[\frac{16}{3} + \frac{1036\,e_t^2}{15}+\frac{1963\,e_t^4}{15}+ \frac{1441\,e_t^6}{30}+ \frac{141\,e_t^8}{80}\Bigg]
\nonumber\\
& \quad
- \frac{\tilde{\mu}_-^2}{(1-e_t^2)^{13/2}}\Bigg[
\frac{42}{5} - \frac{152\nu}{9} + \bigg(\frac{2366}{15} - \frac{10474\nu}{45}\bigg)e_t^2 + \bigg(\frac{1917}{5} - \frac{41339\nu}{90}\bigg)e_t^4 
\nonumber\\
& \hspace{4cm} + \bigg(\frac{9349}{60} - \frac{10241\nu}{60}\bigg)e_t^6 + \bigg(\frac{1773}{320} - \frac{991\nu}{160}\bigg)e_t^8\Bigg]
\,.
\end{align}
\end{subequations}
Quite naturally, the Newtonian factors of Eqs.~\eqref{eq_detdt_split} and~\eqref{eq_dardt_split} are nothing but the well-known Peters' formulas~\cite{Peters:1964zz}. 
As for the 1PN point-particle sectors, we have a full agreement with Eqs.~(6.17) and~(6.19) of~\cite{Arun:2009mc}.

As for the evolution of the orbital period, it comes
\begin{equation}\label{eq_dxdt_split}
\left\langle \frac{\dd x}{\dd t}\right\rangle = \frac{c^3\,\nu}{G m}\,x^5\,\Bigg[\mathcal{X}_\text{N} + x\,\mathcal{X}_{1\text{PN}} + \aem\,x^2\bigg(\mathcal{X}_\text{mag,LO}+x\,\mathcal{X}_\text{mag,NLO}\bigg)\Bigg]\,,
\end{equation}
with
\begin{subequations}
\begin{align}
\mathcal{X}_\text{N}
&
=
\frac{1}{(1-e_t^2)^{7/2}}\Bigg[\frac{64}{5} + \frac{584\,e_t^2}{15} + \frac{74\,e_t^4}{15}\Bigg]\,,\\
\mathcal{X}_{1\text{PN}} 
&
=
\frac{1}{(1-e_t^2)^{9/2}}\Bigg[-\frac{2972}{105} - \frac{176\nu}{5} + \bigg(\frac{1462}{7} - 380\nu\bigg)e_t^2
\nonumber\\
& \hspace{3cm} + \bigg(\frac{12217}{30} - \frac{1687\nu}{5}\bigg)e_t^4 + \bigg(\frac{11717}{420} - \frac{296\nu}{15}\bigg)e_t^6\Bigg]\,,\\
\mathcal{X}_\text{mag,LO}
&
=
\frac{\tilde{\mu}_+^2+\delta\,\tilde{\mu}_+\tilde{\mu}_--\nu\tilde{\mu}_-^2}{(1-e_t^2)^{11/2}}\Bigg[-128 - \frac{10768\,e_t^2}{15} - \frac{7472\,e_t^4}{15} - \frac{118\,e_t^6}{5}\Bigg]
\nonumber\\
&\quad
+ \frac{\tilde{\mu}_-^2}{(1-e_t^2)^{11/2}}\bigg[\frac{8}{15} + \frac{76\,e_t^2}{15}+\frac{23\,e_t^4}{5}+ \frac{3\,e_t^6}{10}\Bigg]
\,,\\
\mathcal{X}_\text{mag,NLO}
&
=\frac{\tilde{\mu}_+^2+\delta\,\tilde{\mu}_+\tilde{\mu}_--\nu\tilde{\mu}_-^2}{(1-e_t^2)^{13/2}}\Bigg[
-\frac{18768}{35} + \frac{2368\nu}{5} - \bigg(6812 - \frac{382784\nu}{45}\bigg)e_t^2 - \bigg(\frac{366455}{21} - \frac{828794\nu}{45}\bigg)e_t^4 
\nonumber\\
& \hspace{5cm} - \bigg(\frac{203953}{30} - \frac{306497\nu}{45}\bigg)e_t^6 - \bigg(\frac{20513}{112} - \frac{3133\nu}{15}\bigg)e_t^8\Bigg]
\nonumber\\
& \quad
+ \frac{\tilde{\mu}_+^2+\delta\,\tilde{\mu}_+\tilde{\mu}_--\nu\tilde{\mu}_-^2}{(1-e_t^2)^6}\,\frac{2(7-2\nu)(96-4484\,e_t^2-4530\,e_t^4-7\,e_t^6)}{45}
\nonumber\\
& \quad
+ \frac{\delta\,\tilde{\mu}_+\tilde{\mu}_-}{(1-e_t^2)^{13/2}}\Bigg[\frac{16}{3} + \frac{1036\,e_t^2}{15}+\frac{1963\,e_t^4}{15}+ \frac{1441\,e_t^6}{30}+ \frac{141\,e_t^8}{80}\Bigg]
\nonumber\\
& \quad
+ \frac{\tilde{\mu}_-^2}{(1-e_t^2)^{13/2}}\Bigg[
\frac{42}{5} - \frac{152\nu}{9} + \bigg(\frac{2246}{15} - \frac{10474\nu}{45}\bigg)e_t^2 + \bigg(\frac{1857}{5} - \frac{41339\nu}{90}\bigg)e_t^4 
\nonumber\\
& \hspace{4cm} + \bigg(\frac{9289}{60} - \frac{10241\nu}{60}\bigg)e_t^6 + \bigg(\frac{1773}{320} - \frac{991\nu}{160}\bigg)e_t^8\Bigg]
\,,
\end{align}
\end{subequations}
and we recognize the Newtonian sector as the ``enhancement'' factor due to Peters and Mathews~\cite{Peters:1963ux}.

\bibliography{ListeRef_GWmag_phase.bib}

\end{document}